\documentclass{aa}

\usepackage{natbib}
\bibpunct{(}{)}{;}{a}{}{,} 

\usepackage{lscape}

\usepackage{graphicx}

\def\ang{\AA}

\def\halpha{\mbox{H$\alpha$}}

\def\deg{\hbox{$^\circ$}}
\def\sun{\hbox{$\odot$}}

\def\lesssim{\mathrel{\hbox{\rlap{\hbox{\lower4pt\hbox{$\sim$}}}\hbox{$<$}}}}
\def\gtrsim{\mathrel{\hbox{\rlap{\hbox{\lower4pt\hbox{$\sim$}}}\hbox{$>$}}}}
\def\la{\mathrel{\hbox{\rlap{\hbox{\lower4pt\hbox{$\sim$}}}\hbox{$<$}}}}
\def\ga{\mathrel{\hbox{\rlap{\hbox{\lower4pt\hbox{$\sim$}}}\hbox{$>$}}}}

\def\am{\hbox{$^\prime$}}
\def\arcsec{\hbox{$^{\prime\prime}$}}

\def\hr{\hbox{$^{\rm h}$}}
\def\min{\hbox{$^{\rm m}$}}
\def\sec{\hbox{$^{\rm s}$}}

\def\farcs{\hbox{$.\!\!^{\prime\prime}$}}

\def\micron{\hbox{$\mu$m}}

\makeatletter
\def\ion#1#2{#1$\;${\small\rm\@Roman{#2}}\relax}
\def\beam{$\theta_{\rm B}$} 

\newcommand{\figsp}{\hspace{7mm}}

\begin{document}

\title{Large dust particles in disks around T~Tauri stars}

\author{J. Rodmann    \inst{1}, 
	Th. Henning   \inst{1},
	C. J. Chandler\inst{2},
        L. G. Mundy   \inst{3}, \and
        D. J. Wilner  \inst{4}
       }

\offprints{Jens Rodmann, \\ \email{rodmann@mpia.de}}

\institute{Max-Planck-Institut f\"ur Astronomie, 
           K\"onigstuhl 17, D--69117, Heidelberg, Germany
	   \and
           National Radio Astronomy Observatory, P.O. Box O, Socorro, NM 87801, USA
           \and
           Department of Astronomy, University of Maryland, College Park, MD 20742, USA
           \and
           Harvard-Smithsonian Center for Astrophysics, 60 Garden Street,
           Cambridge, MA 02138, USA
          }

\date{Received August 12, 2005; accepted September 14, 2005}

\abstract{We present 7-mm continuum observations of 14 low-mass
  pre-main-sequence stars in the Taurus-Auriga star-forming region obtained
  with the Very Large Array with $\sim$1\farcs5 resolution and $\sim$0.3~mJy 
  rms sensitivity. For 10~objects, the circumstellar emission has been
  spatially resolved. The large outer disk radii derived suggest that the 
  emission at this wavelength is mostly optically thin. The millimetre
  spectral energy distributions are characterised by spectral indices 
  \mbox{$\alpha_{\rm mm}=$~2.3 to 3.2}. After accounting for contributions
  from free-free emission and corrections for optical depth, we determine dust 
  opacity indices 
  $\beta$ in the range \mbox{0.5 to 1.6}, which suggest that millimetre-sized
  dust aggregates are present in the circumstellar disks.
  Four of the sources with $\beta>1$ may be consistent with submicron-sized 
  dust as found in the interstellar medium. Our findings indicate that 
  dust grain growth to millimetre-sized particles is completed 
  within less than 1~Myr for the majority of circumstellar disks.

  






}

\authorrunning{Rodmann et al.}
\titlerunning{Large dust particles in disks around T~Tauri stars}
\maketitle
%
\section{Introduction}
Although dust grains constitute only a minor fraction of the disk
material around young stars ($\sim$1 Myr), they play a pivotal role in the 
complicated multi-stage process of planet formation. The growth of 
submicron-sized dust particles as found in the interstellar medium
(ISM) to larger aggregates by coagulation 
(collisional sticking) is the first essential step towards building 
larger rocky bodies (planetesimals) that may eventually accrete into 
terrestrial planets and planetary cores \citep{beckwith00}.

Statistical analysis of the near-infrared excess of young stellar clusters 
shows that disk emission disappears within a few Myr~\citep{strom89,haisch01}. 
Similarly, sub-mm and mm observations suggest a decrease in the amount of cold
circumstellar dust during the post T-Tauri phase 
\citep{carpenter05}. The decline of dust emission can be interpreted as 
a consequence of gradual mass loss through disk dissipation and/or opacity
changes due to particle growth.
It is important to stress that such studies only trace the evolution 
of the dust disk. The evolution of the gaseous disk, containing 
the bulk of the disk mass, is at present only poorly understood. 

For many years there has been little sound observational evidence for dust
grain growth in disks around pre-main-sequence stars. The analysis of spectral energy 
distributions showed that the sub-mm/mm fluxes of T~Tauri stars
decline more slowly towards longer wavelengths than expected for ISM-sized dust
\citep[e.\,g.][]{beckwith91}. It was tempting to interpret a
shallow spectral slope as an indication of the presence of particles much
larger than in the ISM ($\gg0.1$\thinspace\micron). 
It was soon realised, however, that spatially unresolved disk observations 
cannot be used to distinguish between small, optically thick disks containing 
sub-micron ISM dust and extended, optically thin disks with larger dust
particles. Spatially resolved images are needed to break this parameter 
degeneracy~\citep{koerner95,dutrey96}. 

Recent technical improvements at the Very Large Array (VLA) allowed 
the resolution of the disk around the young star \object{TW~Hya} at 7~mm
and the determination of the dust opacity index \citep{wilner00,calvet02}.
\citet{testi03} resolved the dusty disk around the pre-main-sequence 
star \object{CQ~Tau} and concluded that millimetre- and even centimetre-sized 
dust particles must be present in the outer disk. In a similar study of 
six isolated intermediate-mass (Herbig Ae) stars by \cite{natta04}, dust grain 
growth was inferred for two objects. 

In this paper, we report the results of 7-mm continuum observations of 
14 T~Tauri stars located in the Taurus-Auriga star-forming region, currently 
the largest sample of low-mass pre-main-sequence stars investigated for 
signs of dust grain growth to millimetre/centimetre particle sizes.

\begin{table*}[t]
\caption{Properties of sample stars.}
\begin{center}
\begin{tabular}{l l l c l l r@{.}l r@{.}l l } 
\hline\\[-3mm]
Star                                         &
\multicolumn{2}{c}{Coordinates (J2000.0)}    & 
Spectral                                     &
$T_{\rm eff}^\mathrm{(b)}$                   & 
$M_{\star}^\mathrm{(c)}$                     & 
\multicolumn{2}{c}{$L_{\star}^\mathrm{(b)}$} & 
\multicolumn{2}{c}{Age$^\mathrm{(c)}$}       & 
Location$^\mathrm{(d)}$ \\
                                                                   &     
\multicolumn{1}{c}{$\alpha \left(\hr\ \min\ \sec \right)$}         &
\multicolumn{1}{c}{$\delta \left(\degr\ \arcmin\ \arcsec \right)$} &
type$^\mathrm{(a)}$                                                  & 
(K)                                                                & 
$({\rm M}_{\sun})$                                                 & 
\multicolumn{2}{c}{$({\rm L}_{\sun})$}                             &
\multicolumn{2}{c}{(Myr)} \smallskip\\ 
\hline\hline\\[-2.5mm]
\object{CY~Tau}   &  4 17 33.74 &   +28 20 46.5 &  M1	& 3720     & 0.48	&  0&47     & 0&7		      &  L\,1495  \\
\object{RY~Tau}   &  4 21 57.42 &   +28 26 35.5 &  K1	& 5080     & 1.69	&  7&60     & 0&2		      &  L\,1495  \\
\object{FT~Tau}   &  4 23 39.19 &   +24 56 14.2 &  cont.& 3890$^\mathrm{(c)}$     & \multicolumn{1}{c}{---}	&  \multicolumn{2}{c}{---}   &\multicolumn{2}{c}{---}  &  B\,217   \\
\object{DG~Tau~B} &  4 27 02.56 &   +26 05 30.4 &  ---	&  \multicolumn{1}{c}{---}	   &\multicolumn{1}{c}{---}         &  \multicolumn{2}{c}{---} &\multicolumn{2}{c}{---} &  B\,217   \\
\object{DG~Tau}   &  4 27 04.69 &   +26 06 16.1 &  M	& 3890$^\mathrm{(c)}$     & 0.56	&  1&7$^\mathrm{(c)}$  & 0&3		      &  B\,217   \\
\object{HL~Tau}   &  4 31 38.41  &  +18 13 57.6 &  K7--M2 & 4060   & 0.55	&  0&9$^\mathrm{(c)}$  & 1&0		      &  L\,1551  \\
\object{GG~Tau}   &  4 32 30.39 &   +17 31 40.1 &  K7	& 4060     & 0.65	&  1&50     & 0&3		      &  B\,217   \\
\object{UZ~Tau~E} &  4 32 43.10 &   +25 52 31.0 &  M1--3  & 3680   & 0.44	& 2&94$^\mathrm{(c)}$  &$<$0&1		      &  B\,217   \\
\object{DL~Tau}   &  4 33 39.09 &   +25 20 37.9 &  K7	& 4060     & 0.56	& 0&77$^\mathrm{(c)}$  & 1&2		      &  TMC\,2 \\
\object{DM~Tau}   &  4 33 48.74  &  +18 10 10.0 &  M1	& 3720     &0.62        & 0&25     & 3&2		      &  L\,1551  \\
\object{CI~Tau}   &  4 33 52.03 &   +22 50 30.2 &  K7	& 4060     & 0.70       &0&87       & 0&8		      &  L\,1536  \\
\object{DO~Tau}   &  4 38 28.59 &   +26 10 49.4 &  M0	& 3850     &0.72        &  1&20      & 0&6		      &  TMC\,1 \\
\object{LkCa~15}  &  4 39 17.81 &   +22 21 03.6 &  K5	& 4350
&\multicolumn{1}{c}{---}	&  0&74  & 1&0$^\mathrm{(e)}$		      &  L\,1536  \\
\object{GM~Aur}   &  4 55 10.98 &   +30 21 59.5  &  K3   & 4730     &0.72	&  0&83     & 1&8		      & L\,1517  \smallskip \\
\hline
\end{tabular}
\label{sample}
\end{center}
\vspace*{-2mm}
References ---
(a)~\citet{herbig88}; 
(b)~\citet{kenyon95}; 
(c)~\citet{beckwith90}; 
(d)~\citet{leinert93}; 
(e)~\citet{duvert00}.
\end{table*}

\section{Observations}

\subsection{Sample of objects}
Our sample consists of 14 low-mass pre-main-sequence stars located towards the
Taurus-Auriga star formation region (Table~\ref{sample}). All sample 
stars are catalogued as classical T~Tauri stars, based on the equivalent 
width of their \halpha\ emission line $\gtrsim$10~\ang\ \citep{herbig88}. 
They are classified as Class II young stellar objects, as derived from the 
slope of their infrared spectral 
energy distributions, \mbox{$\alpha_{\rm IR} = {\rm d}\log ( \lambda
  F_{\lambda} )/ {\rm d}\log\lambda $}, with \mbox{$-2 \lesssim \alpha \lesssim 0$} 
\citep{adams87,kenyon95,hartmann02}. 

Observations of the millimetre continuum emission have confirmed the presence
of circumstellar disks around these T~Tauri stars, with disk masses ranging
from \mbox{0.02 to 0.7 ${\rm M}_{\sun}$} \citep{beckwith90,dutrey96}. 
For the majority of the stars in our sample, emission from circumstellar 
CO gas has been detected, often tracing a disk in Keplerian rotation
\citep{koerner93,dutrey94,koerner95b,handa95,dutrey96,jensen96,mitchell97,
duvert00,najita03}.

The stellar ages and masses have been determined by fitting pre-main-sequence
isochrones to loci of the stars in the Hertzsprung-Russell diagram; they range from 
\mbox{0.1 to 3.2 Myr} and from \mbox{0.4 to 1.7 ${\rm M}_{\sun}$}, respectively 
\citep{beckwith90,duvert00}. The uncertainties of the age determinations are 
considerable and preclude a meaningful correlation with the opacity indices $\beta$.

The average distance to the Taurus-Auriga molecular cloud as found in the 
literature varies between 135~pc \citep{cernicharo85} and 160~pc \citep{strom89}. 
We adopt the standard value of 140~pc \citep{elias78,kenyon94,loinard05} to 
convert angular scales into physical sizes.



%
\subsection{7-mm continuum observations}
The NRAO\footnote{The National Radio Astronomy Observatory is 
operated for the National Science Foundation by Associated Universities, Inc., 
under a cooperative agreement.} Very Large Array was used to observe the 
dust continuum emission of 14 T~Tauri stars at 43.34~GHz (7~mm). The total 
receiver bandwidth was 100~MHz, divided into two 50~MHz bands centred at 
43.3149~GHz and 43.3649~GHz. 
The observations were carried out in D configuration on 22--23~March~2003 and 
again on 12~May~2003. 
The 27 VLA antennas provided \mbox{$27\times\,(27-1)/2=351$} baselines 
from 35~m to 1.03~km, corresponding to 5--150~k$\lambda$ at 7~mm. 
The May observations took place during array reconfiguration (D to A); 
only 26 antennas were available. 


Phase calibration was accomplished by sandwiching the on-source observations 
between pointings of nearby secondary calibrators (radio sources J0431+2037, 
J0403+2600, J0443+3441). The typical source/phase calibrator cycle time was 
5--10~minutes. 
Absolute flux calibration was obtained from observation 
of the quasar J0542+4951 (3C\,147) which is assumed to have 
a flux density of 0.91~Jy. The estimated uncertainty in the absolute flux
calibration is \mbox{$\sim$10\%}. Reference pointing measurements were made 
approximately every hour. 


The $(u,v)$~data sets for the three observing dates were merged using the
$\mathcal{AIPS}$ task \texttt{DBCON}. CLEANed images were obtained with 
\texttt{IMAGR} using a \texttt{ROBUST} parameter of 0, which is 
intermediate between natural and uniform weighting of the visibility
data and optimises for spatial resolution and sensitivity. Three weak
sources (\object{DM~Tau}, \object{CI Tau}, and \object{LkCa~15}) were 
CLEANed using natural weighting (\texttt{ROBUST}=5) to increase
signal-to-noise ratio and allow secure detection. The extended emission
of \object{GG~Tau} is also best imaged using natural weighting. 
Figure~\ref{7mm_maps} shows the corresponding contour plots for the 7-mm 
images.

\newcommand{\vs}{0.1875}
\onecolumn
\begin{figure}[Ht]
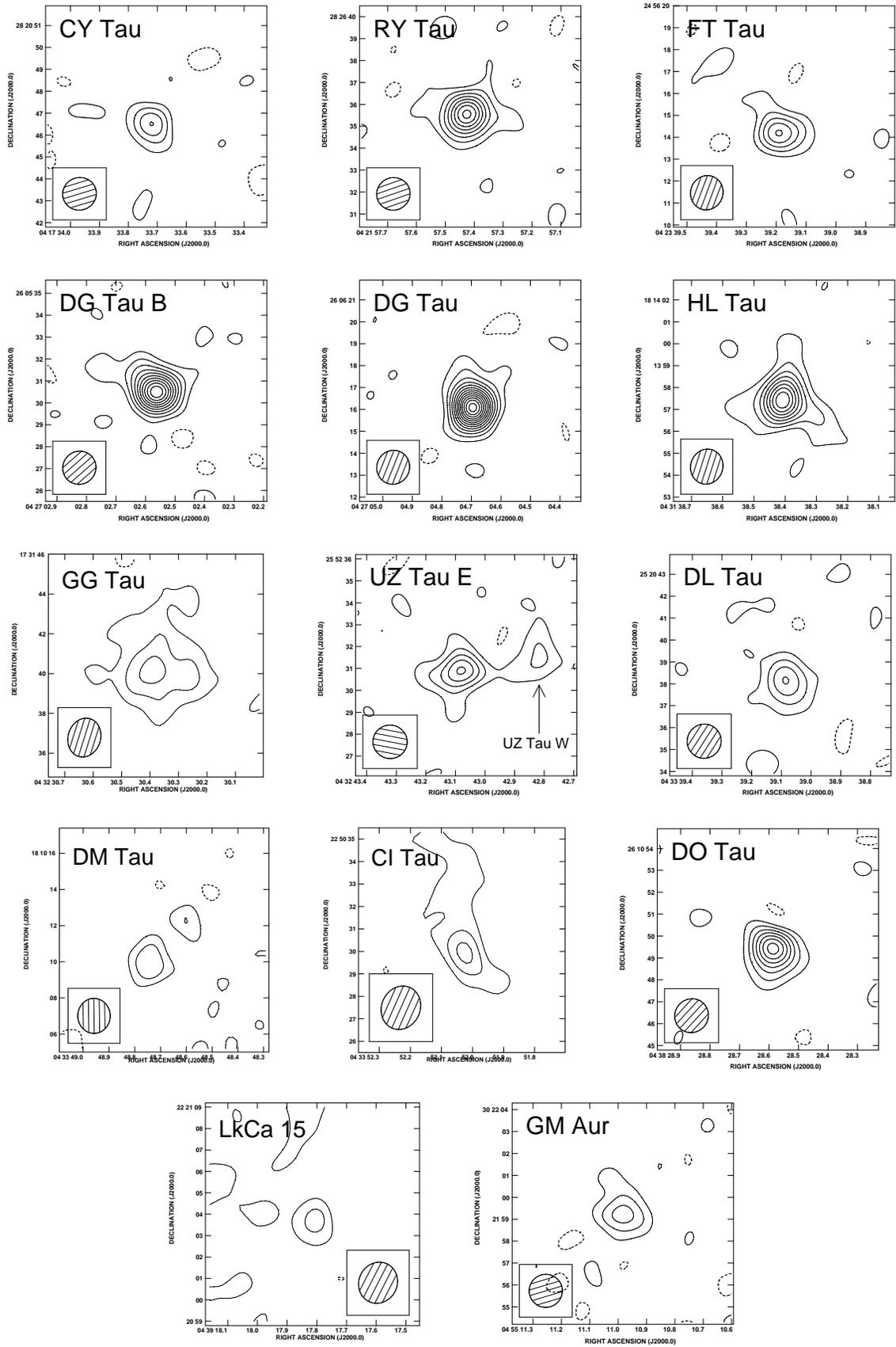
 
\centering 

\if@referee
\includegraphics[width=0.25\textwidth]{CYTAU_7mm.ps}\figsp
\includegraphics[width=0.25\textwidth]{RYTAU_7mm.ps}\figsp 
\includegraphics[width=0.25\textwidth]{FTTAU_7mm.ps}\figsp 
\includegraphics[width=0.25\textwidth]{DGTAUB_7mm.ps}\figsp
\includegraphics[width=0.25\textwidth]{DGTAU_7mm.ps}\figsp
\includegraphics[width=0.25\textwidth]{HLTAU_7mm.ps}\figsp
\includegraphics[width=0.25\textwidth]{GGTAU_7mm.ps}\figsp
\includegraphics[width=0.25\textwidth]{UZTAU_7mm.ps}\figsp
\includegraphics[width=0.25\textwidth]{DLTAU_7mm.ps}\figsp
\includegraphics[width=0.25\textwidth]{DMTAU_7mm.ps}\figsp
\includegraphics[width=0.25\textwidth]{CITAU_7mm.ps}\figsp
\includegraphics[width=0.25\textwidth]{DOTAU_7mm.ps}\figsp
\includegraphics[width=0.25\textwidth]{LKCA15_7mm.ps}\figsp
\includegraphics[width=0.25\textwidth]{GMAUR_7mm.ps}\figsp
\else
\includegraphics[height=\vs\textheight]{CYTAU_7mm.ps}\figsp
\includegraphics[height=\vs\textheight]{RYTAU_7mm.ps}\figsp 
\includegraphics[height=\vs\textheight]{FTTAU_7mm.ps}\figsp 
\includegraphics[height=\vs\textheight]{DGTAUB_7mm.ps}\figsp
\includegraphics[height=\vs\textheight]{DGTAU_7mm.ps}\figsp
\includegraphics[height=\vs\textheight]{HLTAU_7mm.ps}\figsp
\includegraphics[height=\vs\textheight]{GGTAU_7mm.ps}\figsp
\includegraphics[height=\vs\textheight]{UZTAU_7mm.ps}\figsp
\includegraphics[height=\vs\textheight]{DLTAU_7mm.ps}\figsp
\includegraphics[height=\vs\textheight]{DMTAU_7mm.ps}\figsp
\includegraphics[height=\vs\textheight]{CITAU_7mm.ps}\figsp
\includegraphics[height=\vs\textheight]{DOTAU_7mm.ps}\figsp
\includegraphics[height=\vs\textheight]{LKCA15_7mm.ps}\figsp
\includegraphics[height=\vs\textheight]{GMAUR_7mm.ps}\figsp
\fi
\caption{VLA D-configuration images of the $\lambda=7$~mm continuum emission. 
Contour levels are drawn in steps of the corresponding 
2$\sigma$ rms noise level; the average 1$\sigma$ rms noise level is 
$\sim$0.15~mJy/beam. The synthesized beam is shown in the bottom left 
corner of each panel; the average beam size is  $\sim$1\farcs5. 
} 
\label{7mm_maps} 
\end{figure}
\newcommand{\figsz}{0.31}
\begin{figure}[Ht] 
\centering 
\includegraphics[width=\figsz\textwidth]{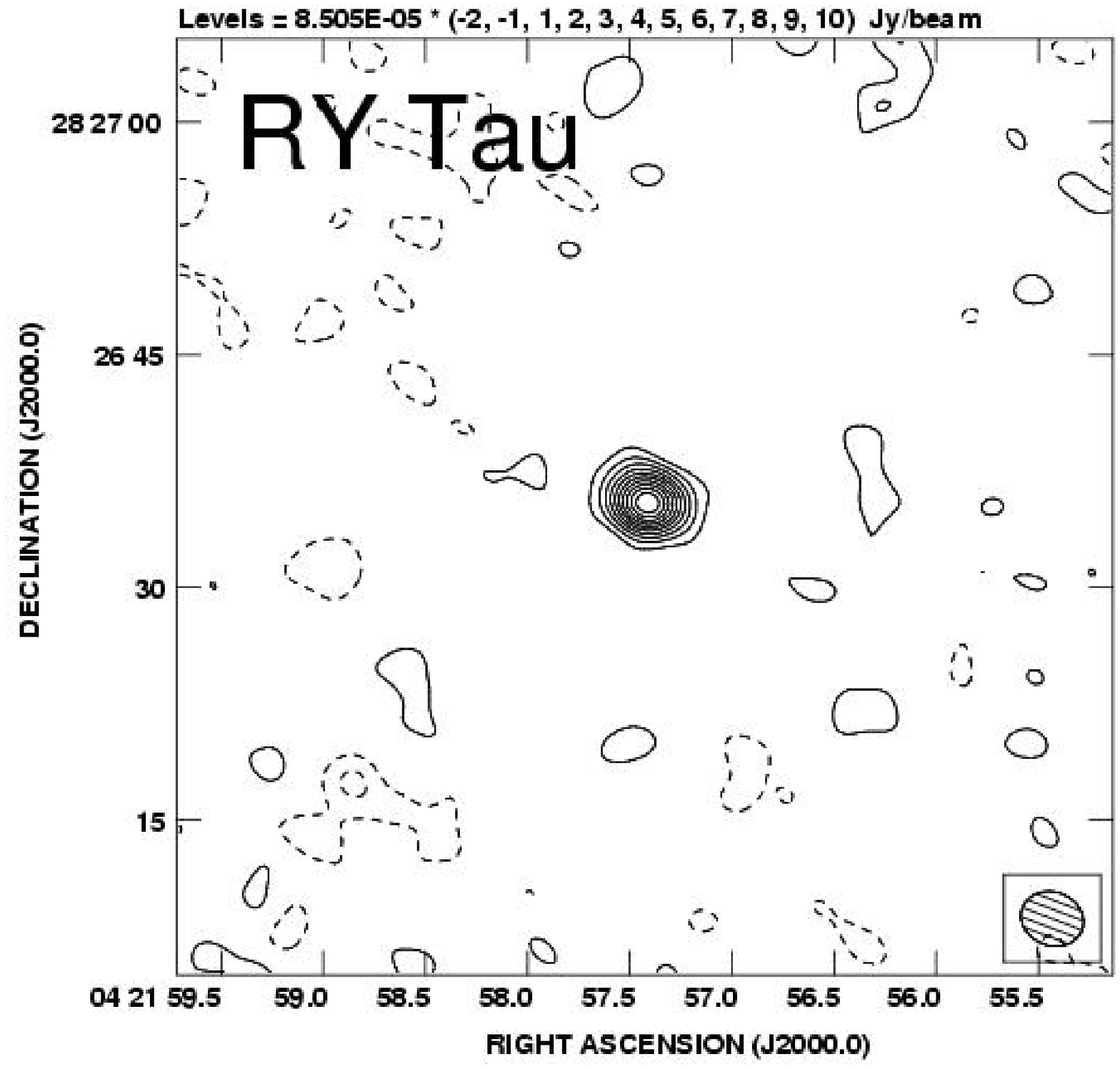}
\includegraphics[width=\figsz\textwidth]{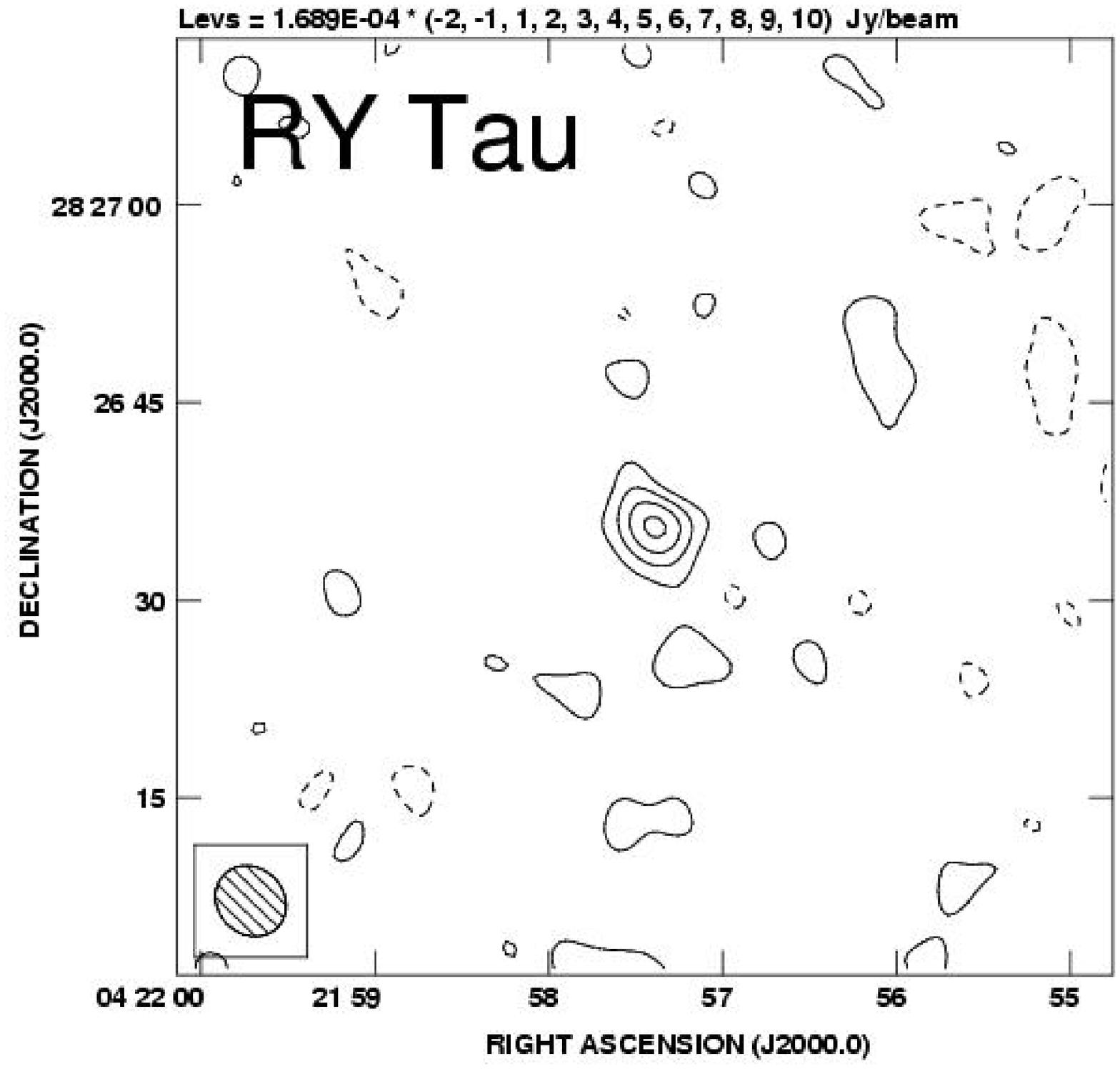}
\includegraphics[width=\figsz\textwidth]{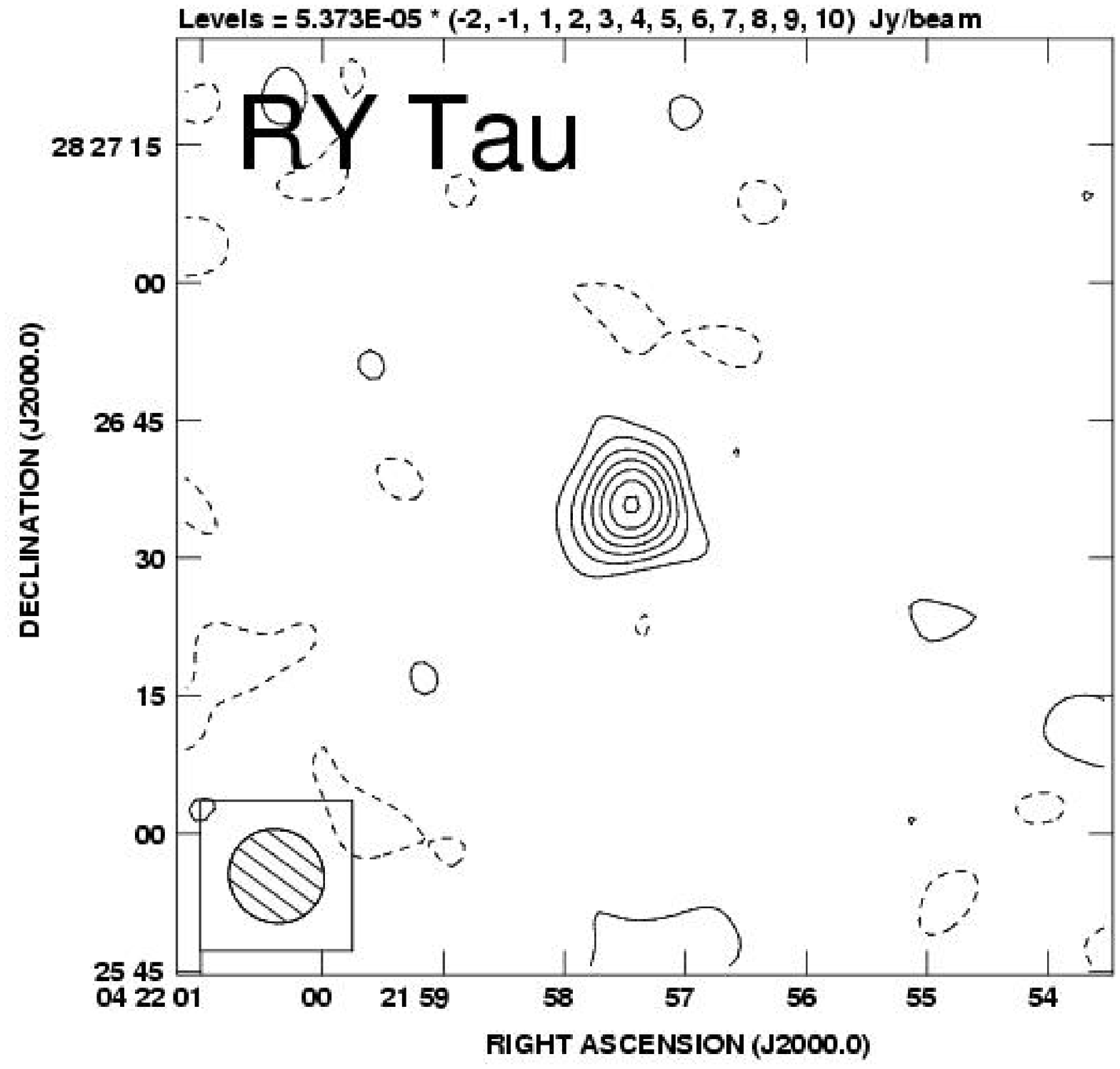}
\includegraphics[width=\figsz\textwidth]{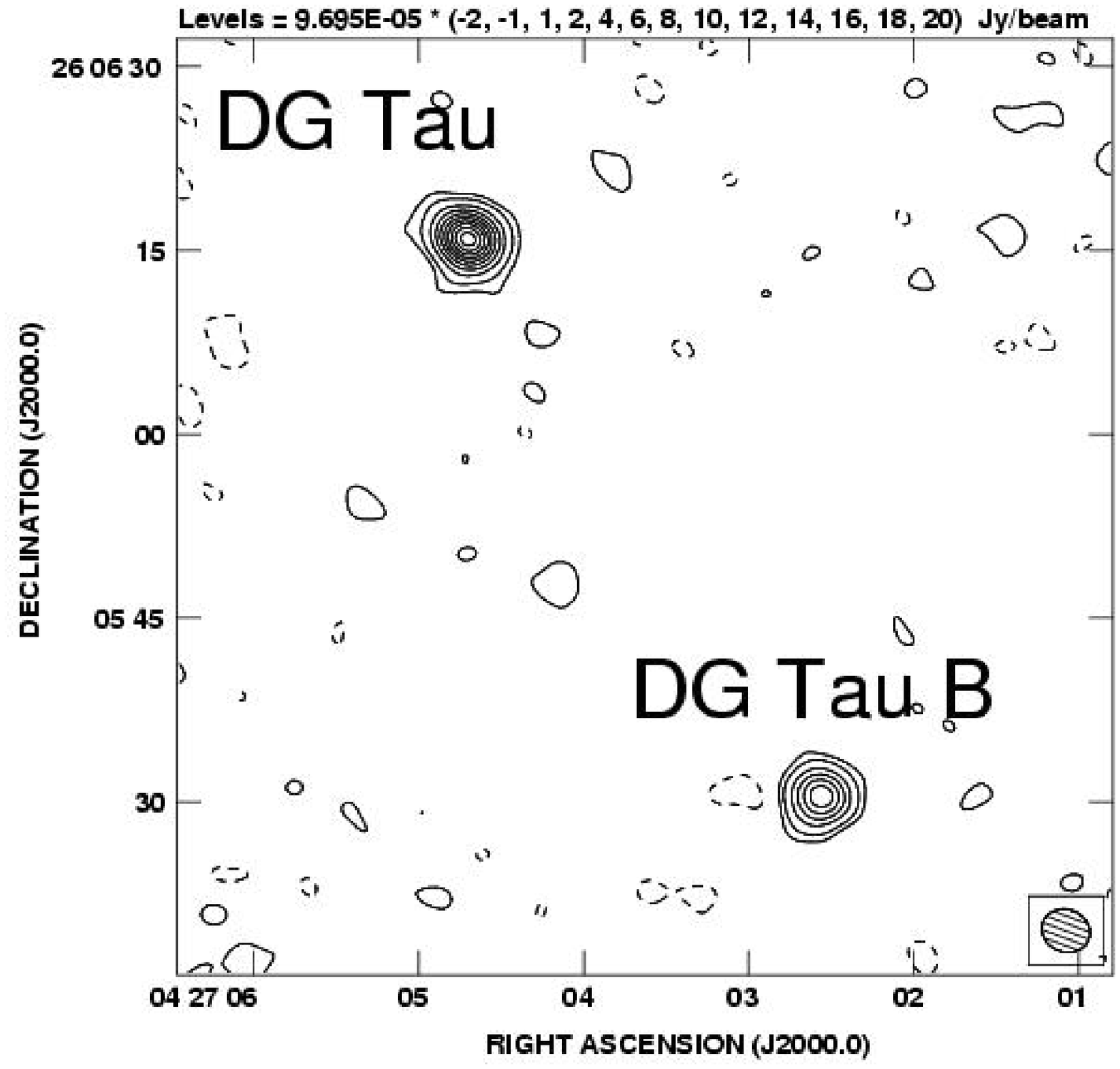} 
\includegraphics[width=\figsz\textwidth]{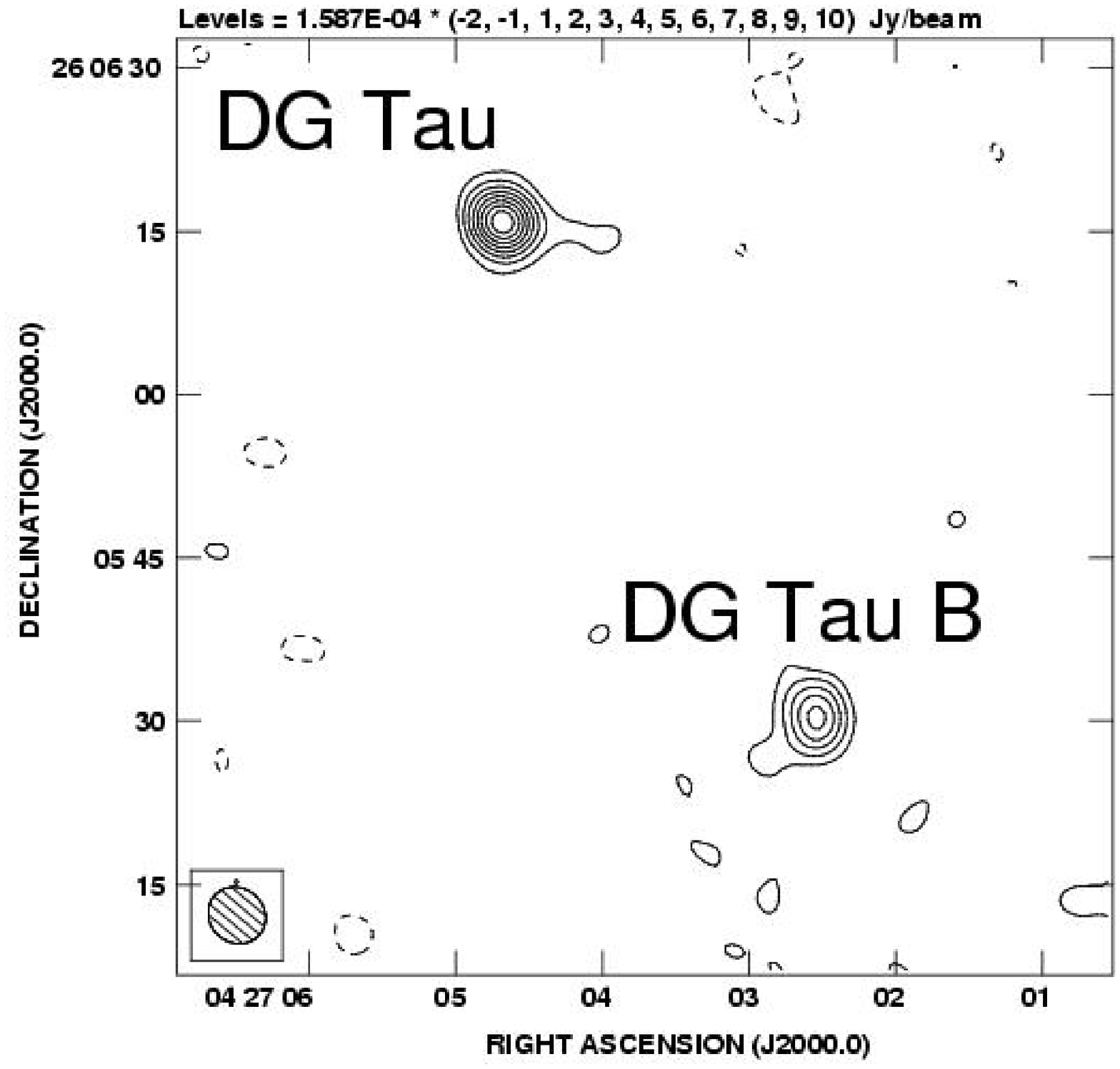} 
\includegraphics[width=\figsz\textwidth]{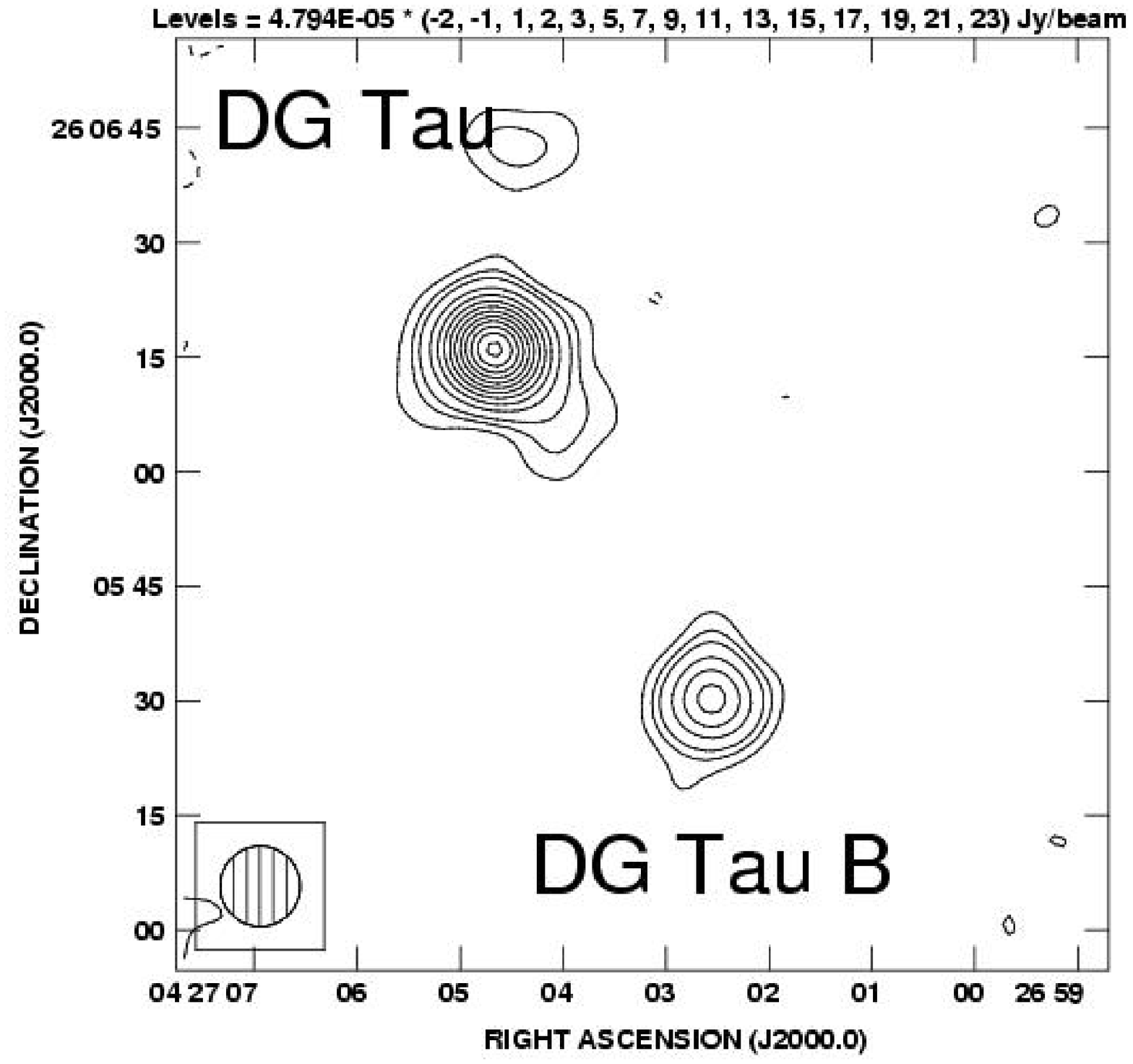} 
\includegraphics[width=\figsz\textwidth]{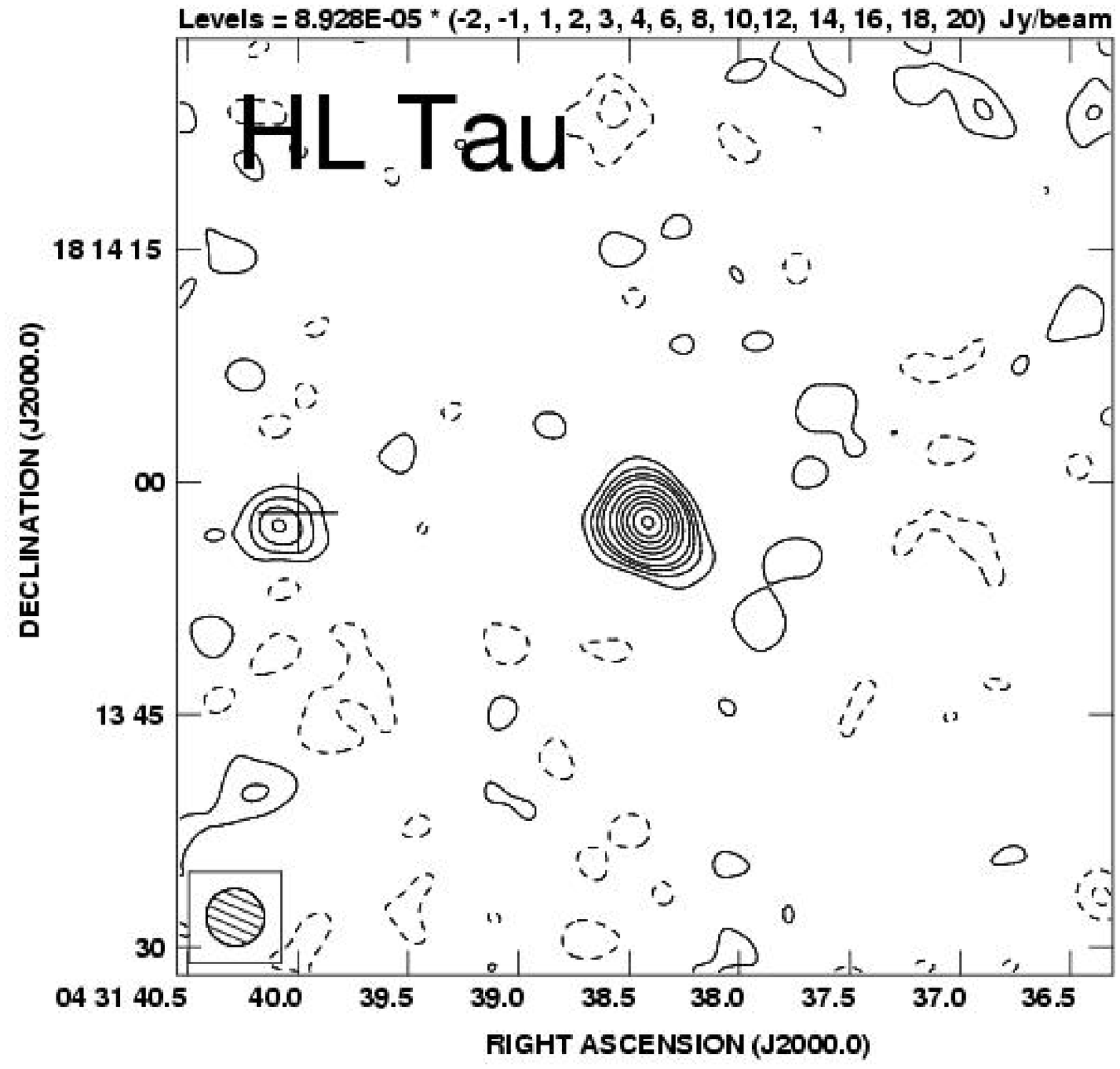}
\includegraphics[width=\figsz\textwidth]{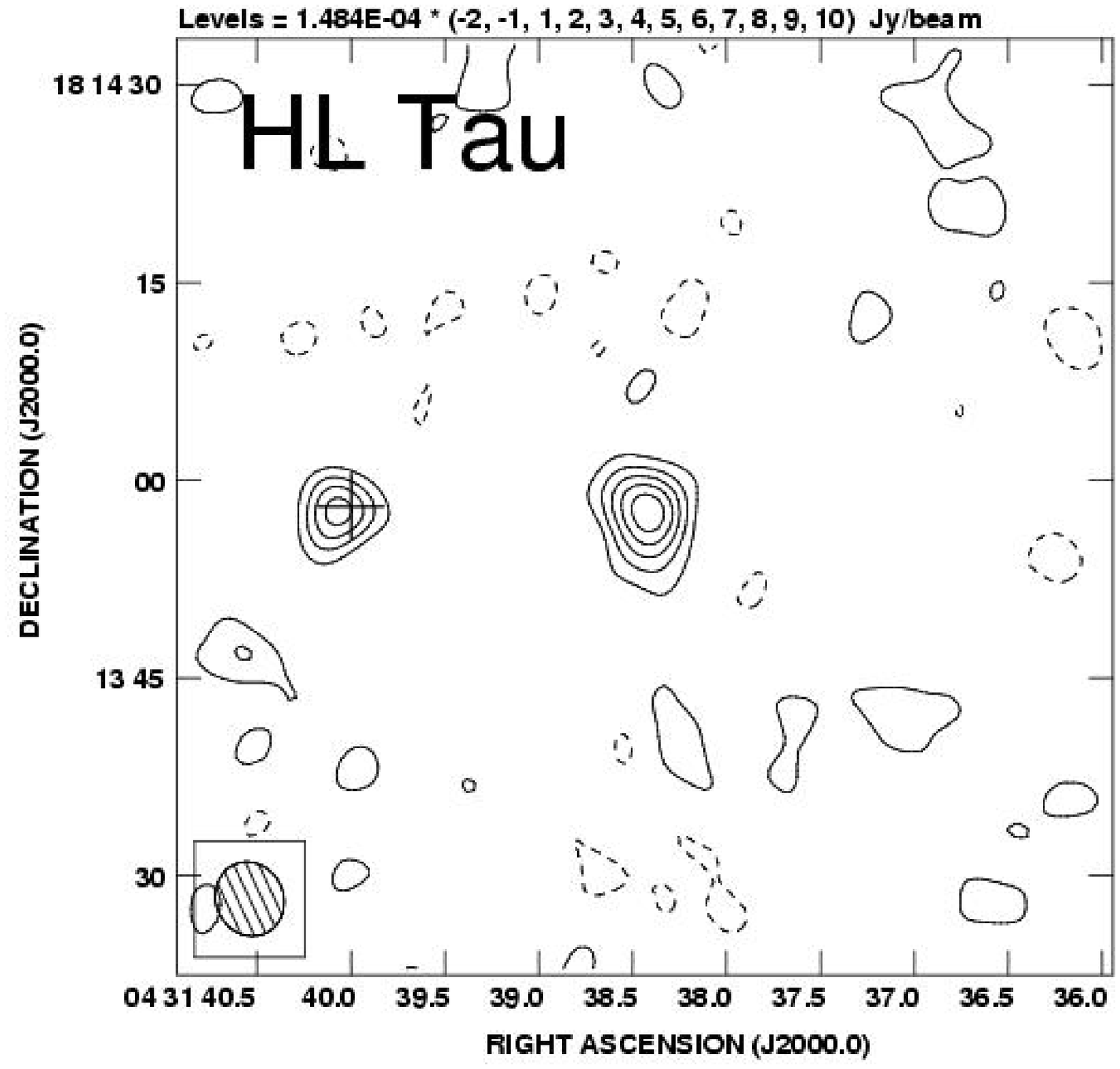}
\includegraphics[width=\figsz\textwidth]{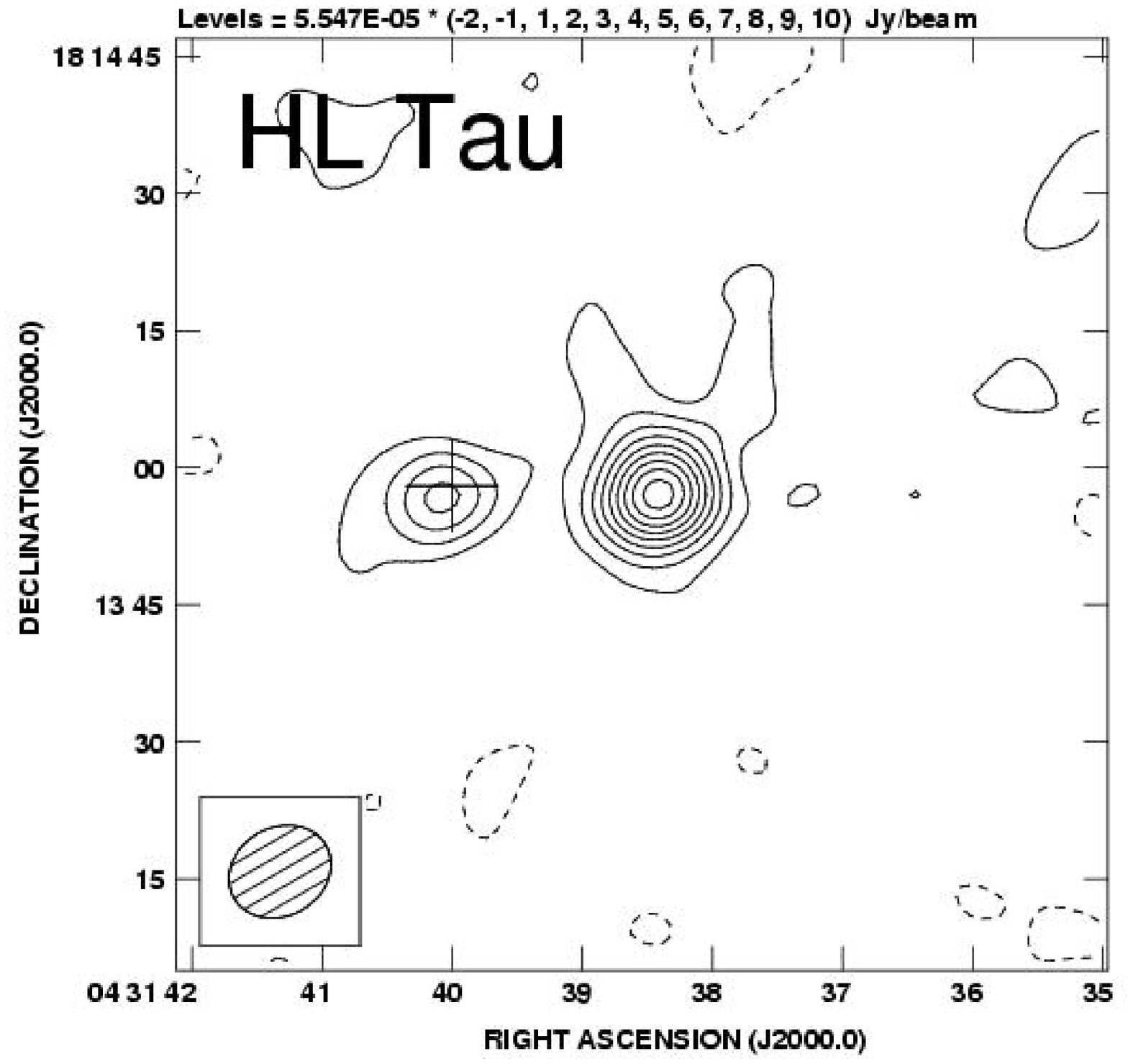}
\includegraphics[width=\figsz\textwidth]{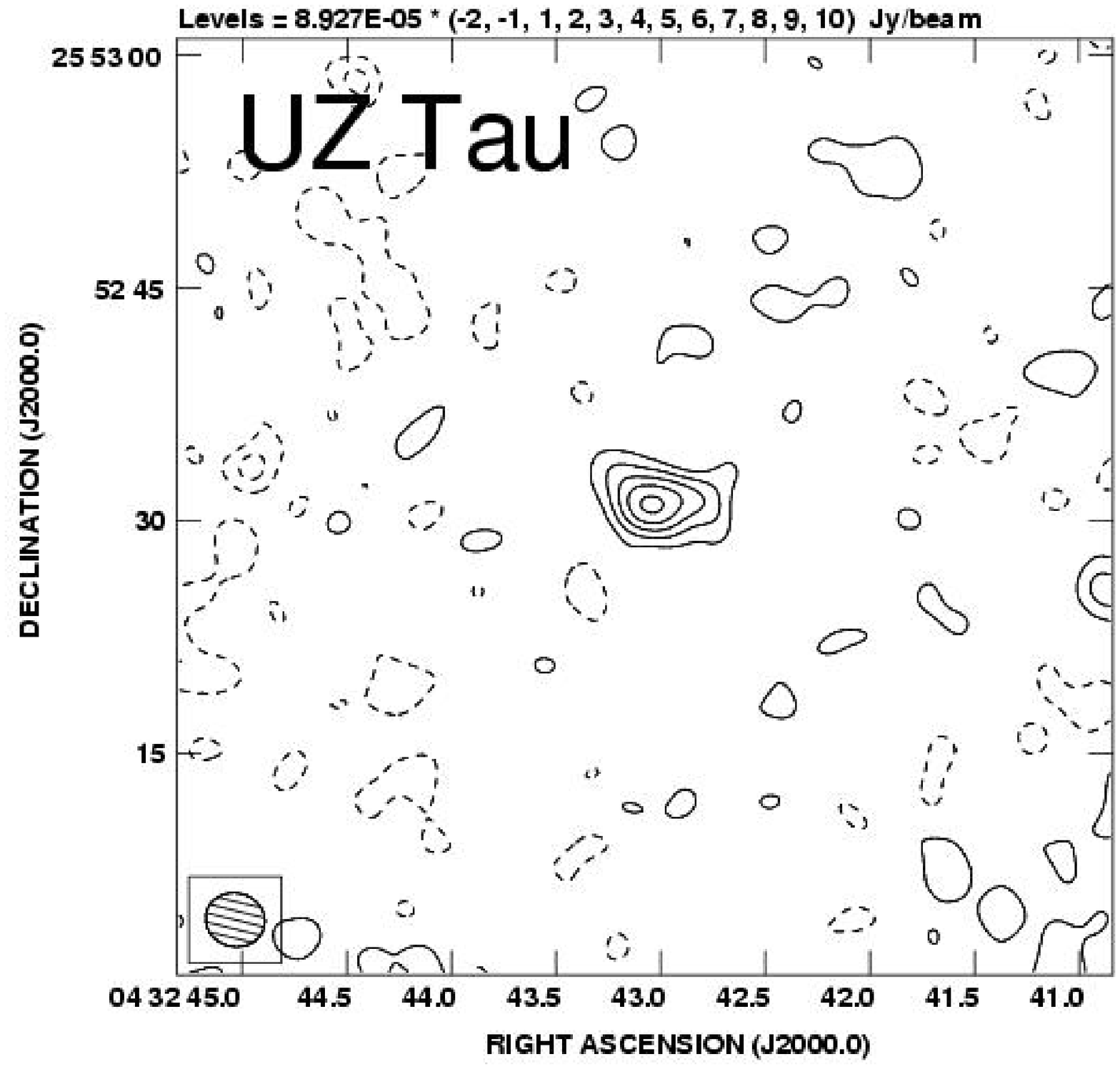}
\includegraphics[width=\figsz\textwidth]{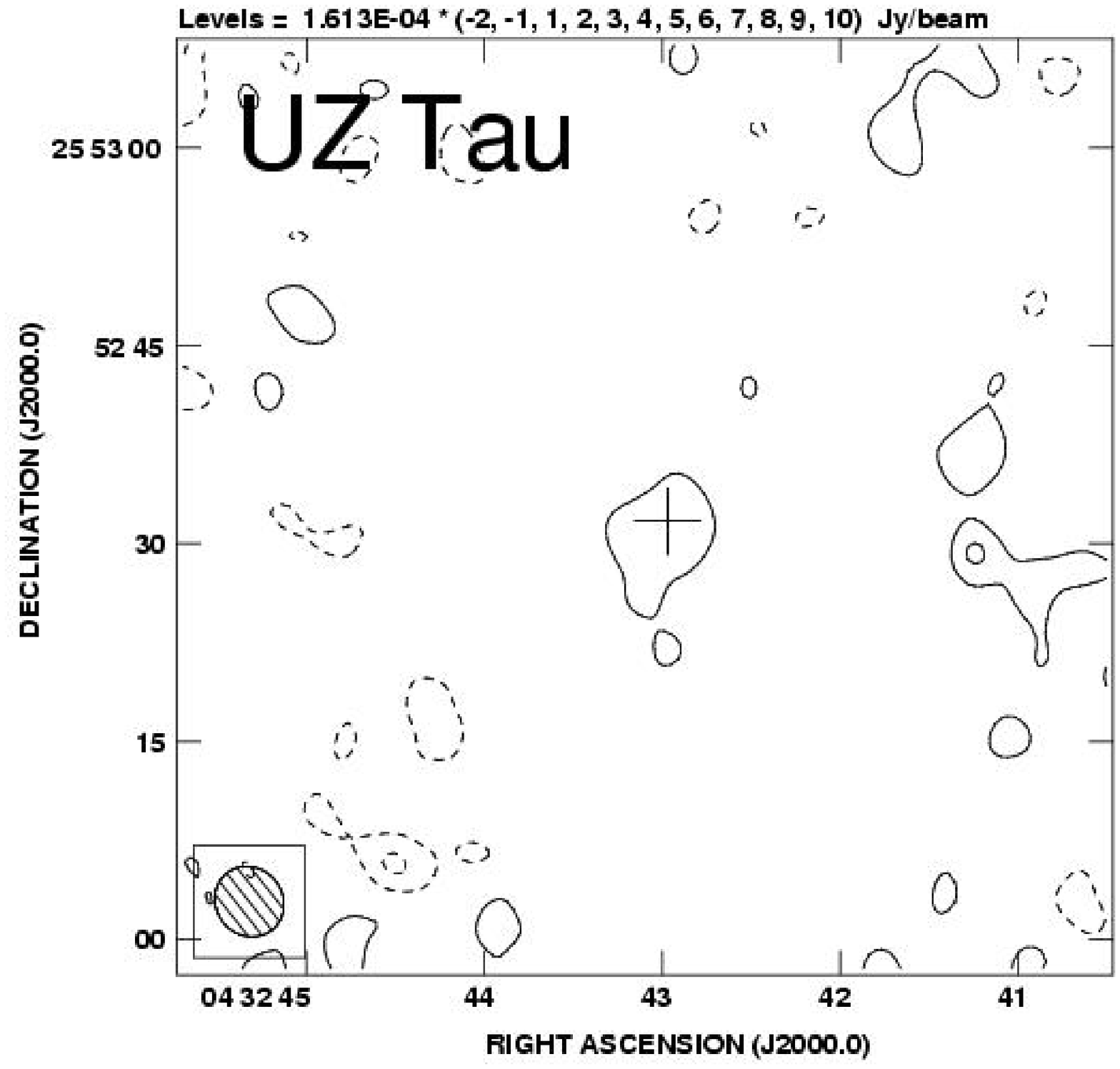}
\includegraphics[width=\figsz\textwidth]{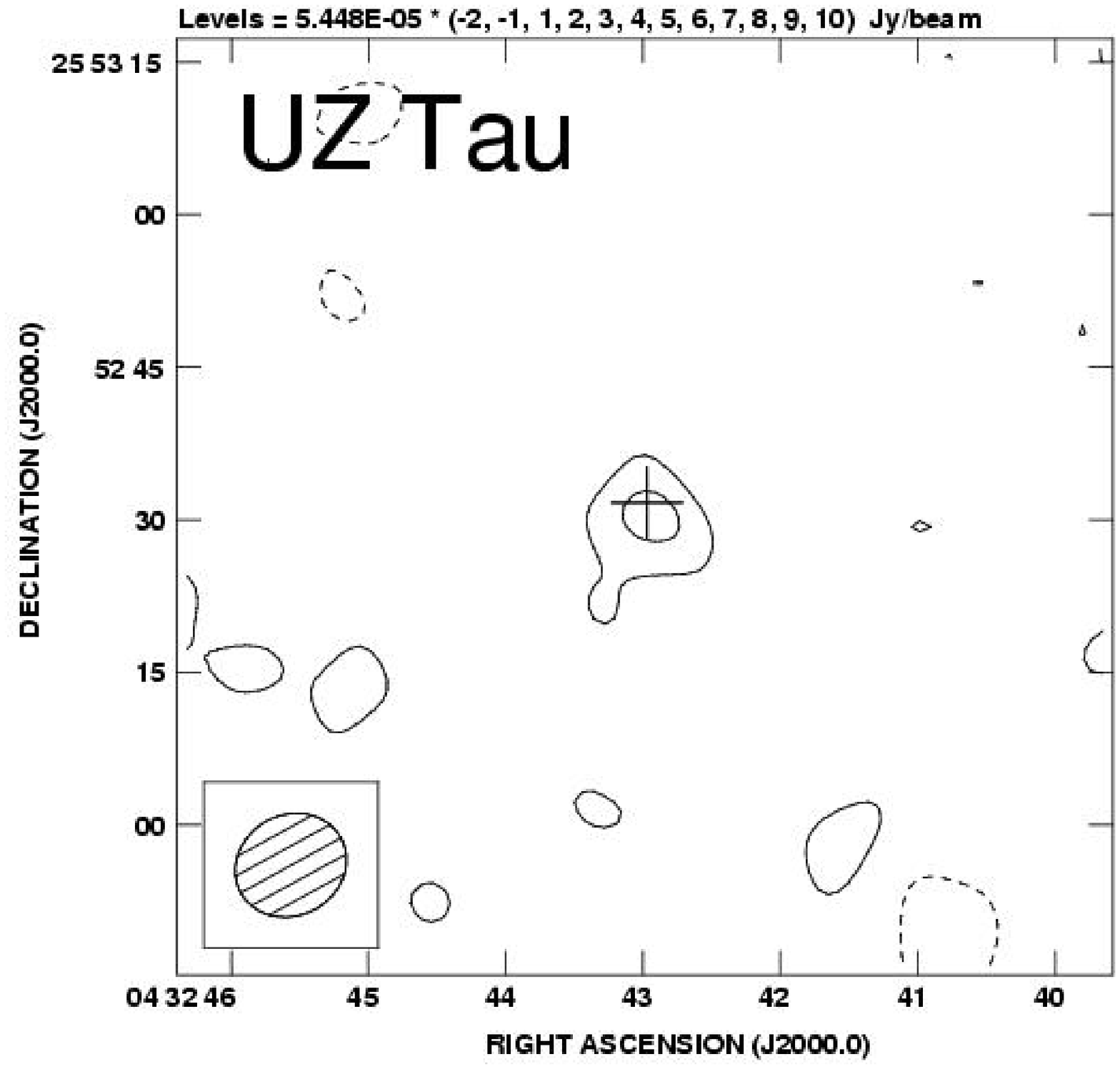}
\caption{VLA D-configuration images at $\lambda=1.3$~cm (left column), 
$\lambda=2.0$~cm (middle column), and $\lambda=3.6$~cm (right column) for 
\object{RY~Tau}, \object{DG~Tau} and \object{DG~Tau~B}, \object{HL~Tau}, 
and \object{UZ~Tau}. Contour levels are given for each panel separately 
in multiples of 2$\sigma$ rms noise level. The average 1$\sigma$ rms noise 
levels are approximately 50, 80, and 30~$\mu$Jy/beam at 1.3, 2.0, and 3.6~cm,
respectively. About 30\arcsec\ to the east of \object{HL~Tau} lies
\object{XZ~Tau} (indicated by crosshairs).
}
\label{cm_maps} 
\end{figure}
%
\if@referee
  \onecolumn
\else
  \twocolumn
\fi
%

%
%
\begin{table*}[t]
\begin{center}
\caption{Seeing-corrected 7-mm flux densities, spatial resolutions, 
and estimated source sizes. The stated uncertainties for the flux
densities do not contain the uncertainty of the absolute flux calibration.} 
\label{7mm_results}
\begin{tabular}{l r@{\,\,$\pm$\,\,}l  c r@{\,\,$\times$\,\,}l r } 
\hline\\[-3mm]
Source                                 & 
\multicolumn{2}{c}{Integrated flux}    &  
Disk resolved?                         &
\multicolumn{2}{c}{Deconvolved  source size}   &
\multicolumn{1}{c}{P.A.}                             \\ 
                                       & 
\multicolumn{2}{c}{(mJy)}              & 
                                       & 
\multicolumn{2}{c}{(AU)}               &
\multicolumn{1}{c}{(\deg)}  
\smallskip \\ \hline\hline\\[-2.5mm] 
%
\vspace*{1mm}
\object{CY Tau}	  &  1.19 &   0.26 &  no        &\multicolumn{2}{c}{$\!\!\!$---}& \multicolumn{1}{c}{---}\\
\vspace*{1mm}
\object{RY Tau}	  &  2.97 &   0.29 &  yes       & 180 ($\pm$30)& 40 ($\pm$50)   & 140 $\pm$ 10 \\
\vspace*{1mm}
\object{FT Tau}	  &  1.62 &   0.27 &  partially & 170 ($\pm$50)& ---& 60 $\pm$ 20 \\
\vspace*{1mm}
\object{DG Tau B} &  3.57 &   0.29 &  yes       & 130 ($\pm$30)& 40 ($\pm$50)
& 50 $\pm$ 30 \\
\vspace*{1mm}
\object{DG Tau}	  &  5.30 &   0.37 &  partially & 110 ($\pm$30)& ---
& 10 $\pm$ 10 \\
\vspace*{1mm}
\object{HL Tau}	  &  4.92 &   0.43 &  yes       & 140 ($\pm$40)& 80 ($\pm$40)
& 170 $\pm$ 30\\
\vspace*{1mm}
\object{GG Tau}	  &  3.24 &   0.42 &  ---       & \multicolumn{2}{c}{}          & \\
\vspace*{1mm}
\object{UZ Tau E} &  1.84 &   0.28 &  partially & 120 ($\pm$50)& ---
& 80 $\pm$ 20 \\
\vspace*{1mm}
\object{DL Tau}	  &  1.49 &   0.25 &  yes       & 210 ($\pm$60)& 140
($\pm$100) & 110 $\pm$ 40\\
\vspace*{1mm}
\object{DM Tau}	  &  0.74 &   0.20 &  ---       &  \multicolumn{2}{c}{}         & \\
\vspace*{1mm}
\object{CI Tau}	  &  0.76 &   0.17 &  partially & 220 ($\pm$60)& ---
& 40 $\pm$ 10 \\
\vspace*{1mm}
\object{DO Tau}	  &  2.47 &   0.29 &  partially &  60 ($\pm$50)& ---
& 70 $\pm$ 30 \\
\vspace*{1mm}
\object{LkCa 15}  &  0.44 &   0.17 &  ---       &  \multicolumn{2}{c}{}         & \\
\vspace*{1mm}
\object{GM Aur}	  &  1.05 &   0.23 &  yes       & 200 ($\pm$50)&  60 ($\pm$70)  & 10 $\pm$ 20\\

\hline
\end{tabular}
\end{center}
\end{table*}
\subsection{Centimetre observations}
Five of the sources have poor-quality archival VLA detections at 8~GHz 
\citep{wilner98}. All five sources were reobserved at 
3.6~cm (8.46~GHz), 2.0~cm (14.94~GHz), and 1.3~cm (22.46~GHz).
The total bandwidth was again 100~MHz, with two 50~MHz
channels 50~MHz apart. The 1.3-cm observations were carried out on 17 March
2003; the 2.0- and 3.6-cm observations on 21~March~2003.
All observations were
made in the compact D configuration, covering baselines of \mbox{3--77~k$\lambda$},
\mbox{2--50~k$\lambda$}, and \mbox{1--30~k$\lambda$} in the $(u,v)$-plane at 
1.3, 2.0, and 3.6~cm, respectively.

The flux density scale was set by observations of the calibrator 
J0542+4951 with known flux densities of 1.80, 2.71, and 4.74~Jy at 
1.3, 2.0, and 3.6~cm, respectively. CLEANed images were produced with 
\texttt{IMAGR} and natural weighting 
of the visibilities (\texttt{ROBUST}=5). Finally, primary beam
corrections were performed using the $\mathcal{AIPS}$ task \texttt{LTESS} to
compensate for the non-uniform single-antenna response across the
field-of-view of the entire array. The contour maps are shown in 
Figure~\ref{cm_maps}; note that \object{DG~Tau} 
and \object{DG~Tau~B} appear in the same image.

\subsection{Radio seeing}
Short-term variations in precipitable water vapour lead to changes in the path 
length of an electromagnetic wave travelling through the troposphere, 
causing phase fluctuations when observed by radio interferometers. 
Tropospheric phase noise negatively affects spatial resolution and coherence,
especially at mm wavelengths~\citep{carilli99}.

In order to estimate the amount of smearing introduced by tropospheric phase noise,
we compared self-calibrated and standard-calibrated images of two bright radio
sources. Images of self-calibrated sources are corrected for the radio seeing
on short timescales (3.3~seconds integration time), and deconvolution will 
give the intrinsic source size. The same procedure applied to images obtained from 
ordinary calibration gives the `seeing disk'. By using two test sources, one
unresolved (J0426+2327), the other extended (J0412+2305), we checked the flux 
reduction by phase-noise induced smearing for point sources and extended
sources, respectively.

The analysis of the merged 7-mm data showed that the overall radio seeing was
$\sim$0\farcs5. Inspection of the single-day data revealed that the
seeing on the second day (23~March~2003) was the worst ($\sim$0\farcs7),
while on 22~March and 12~May better water vapour conditions prevailed
(about 0\farcs1 and 0\farcs4, respectively). The integrated 7-mm
fluxes are reduced by 4--5\%, and are within 2\% for each day.

The seeing at 1.3~cm was about 0\farcs5. The peak fluxes might be
reduced by up to 5\%; the integrated fluxes were affected less than 1\%. 
Tropospheric seeing effects at longer centimetre wavelengths are expected
to be negligible.

\section{Results}

\subsection{7-mm fluxes and disk sizes}
\label{7mm}
We detected the continuum emission from cold circumstellar dust at 7 mm
for all 14 T~Tauri stars comprising our sample. Two sources 
(\object{DM~Tau} and \object{LkCa~15}) have signal-to-noise ratios $<\!5$, 
and we do not consider these sources further in our analysis.
The integrated fluxes were measured on the combined maps using the 
$\mathcal{AIPS}$ routine \texttt{IMSTAT}, and corrected for seeing
reduction. In the light of the radio seeing analysis, we decided 
to use the excellent conditions of the first-day data set to determine 
source sizes. The task \texttt{IMFIT} was used to check if the source was
spatially resolved or unresolved by fitting a two-dimensional Gaussian 
function to the emission peak. The deconvolved disk sizes
are of the order 100--200~AU, representing typical dimensions for
protoplanetary disks. The major and minor axes can, in principle, be used to 
derive the inclination angle of the disks. Their uncertainties, however, 
are very large, and the disk orientations are poorly constrained. 
Table~\ref{7mm_results} summarises the results derived from the 7-mm maps. 

The extended dust emission of three sources is of particular interest.
The map of \object{HL~Tau} shows a prominent ``finger'' pointing in south-western 
direction. The measured position angle of \mbox{$235\degr \pm 10\degr$} is 
roughly co-aligned with the counter-jet of \object{HL~Tau} \citep{mundt88, wilner00b}.



We detected extended dust emission and resolved circumstellar material around the 
northern binary (\mbox{$\sim$0\farcs3}\ separation) of the quadruple system 
\object{GG~Tau}~\citep{leinert93}. Previous $^{13}\rm{CO}$ line emission
and 1.4-mm continuum observations demonstrated that the circumbinary disk 
consists of two structural components: an 80-AU wide ring at a radius of 
\mbox{$\sim$220~AU}, and a fainter, more extended disk \citep{dutrey94,guilloteau99}. 
Given the complicated source geometry of \object{GG~Tau} we have not attempted
to deconvolve a simple Gaussian model from the 7-mm map.

The quadruple system \object{UZ~Tau} could be resolved into the bright eastern 
and the (marginally detected) western components, both of which are known 
to harbour a binary system. Speckle and direct imaging observations resolved 
\object{UZ~Tau~W} into a binary system with a projected separation of
\mbox{$\sim$50~AU} \citep{ghez93,simon95}. Radial-velocity measurements
showed \object{UZ~Tau~E} to be a spectroscopic binary with a projected semimajor 
axis of \mbox{$\sim$0.1~AU} \citep{mathieu96,prato02}. We confirm the 
findings of \cite{jensen96}, who noted a substantial reduction of 
millimetre emission around \object{UZ~Tau~W} compared to the tight binary 
\object{UZ~Tau~E}. In the former system, the two components have a separation 
comparable to a typical protoplanetary disk, leading to disk truncation
and therefore smaller and less massive disks. The two components of 
\object{UZ~Tau~E}, on the other hand, apparently do not affect the common 
circumbinary disk in which they reside \citep{mathieu00}.

\subsection{Centimetre fluxes}
Centimetre emission could be detected for all five sources observed at 
1.3, 2.0, and 3.6~cm (Table~\ref{cm_results}). The integrated fluxes 
of \object{UZ~Tau} at 2.0 and 3.6~cm are below the 4-$\sigma$ level. 
The counter-jet of \object{HL~Tau} mentioned in Section~\ref{7mm} can also be 
seen in the corresponding 1.3-cm image. Its position angle has been measured
to $225\degr \pm 20\degr$, in agreement with the value measured from the 7-mm map. 
The \object{UZ~Tau} multiple system is not resolved into its eastern and
western components.

Radio emission from low-mass pre-main-sequence stars is a common phenomenon
\citep{guedel02}. VLA observations of classical T~Tauri stars at radio
wavelengths revealed rising spectral indices and large angular sizes, 
interpreted as signs of wind emission \citep{cohen82a}.
Observational evidence and theoretical arguments favour focussed anisotropic
outflows and collimated jets rather than uniform, isotropic mass flows from 
T~Tauri stars \citep{cohen82b,cohen86}. Wind emission is likely to 
contribute to the 7-mm continuum emission and has to be subtracted from the 
measured 7-mm flux densitites in order to accurately derive dust opacity 
indices from spectral slopes at millimetre wavelengths.


\begin{table}[]
\begin{center}
\caption{Summary of centimetre observations. 1.3-cm values corrected for
seeing-related flux reduction. The uncertainty in the absolute flux density
calibration is not included.} 
\label{cm_results}
\begin{tabular}{l l r@{\,\,$\pm$\,\,}l c } 
\hline\\[-3mm]
Wavelength   &  Source & \multicolumn{2}{l}{Integrated} & Detection \\ 
             &         & \multicolumn{2}{l}{flux (mJy)} & ($\sigma$)
\smallskip \\ \hline\hline\\[-2.5mm]  
%
%
%
1.3 cm       & \object{RY Tau}	  &  0.92 &  0.08  & 12\\
             & \object{DG Tau B}  &  1.23 &  0.09  & 14\\
             & \object{DG Tau}    &  2.17 &  0.11  & 20\\
             & \object{HL Tau}    &  1.63 &  0.11  & 15\\
             & \object{UZ Tau}	  &  0.77 &  0.11  &  7\smallskip \\ \hline\\[-2.5mm] 

2.0 cm       & \object{RY Tau}	  &  0.63 &  0.13  &  5\\
             & \object{DG Tau B}  &  0.80 &  0.13  &  6\\
             & \object{DG Tau}	  &  1.33 &  0.15  &  9\\
             & \object{HL Tau}    &  0.88 &  0.14  &  6\\
             & \object{UZ Tau}	  &  0.48 &  0.14  &  3\smallskip \\ \hline\\[-2.5mm] 

3.6 cm       & \object{RY Tau}	  &  0.31 &  0.04  &  8\\
             & \object{DG Tau B}  &  0.46 &  0.05  &  9\\
             & \object{DG Tau}	  &  1.27 &  0.05  & 25\\
             & \object{HL Tau}    &  0.57 &  0.05  & 11\\
             & \object{UZ Tau}	  &  0.10 &  0.03  &  3\smallskip \\ \hline
\end{tabular}
\end{center}
\end{table}
\subsection{Free-free emission}
%
\begin{figure*}
  \centering
  \includegraphics[width=0.45\textwidth]{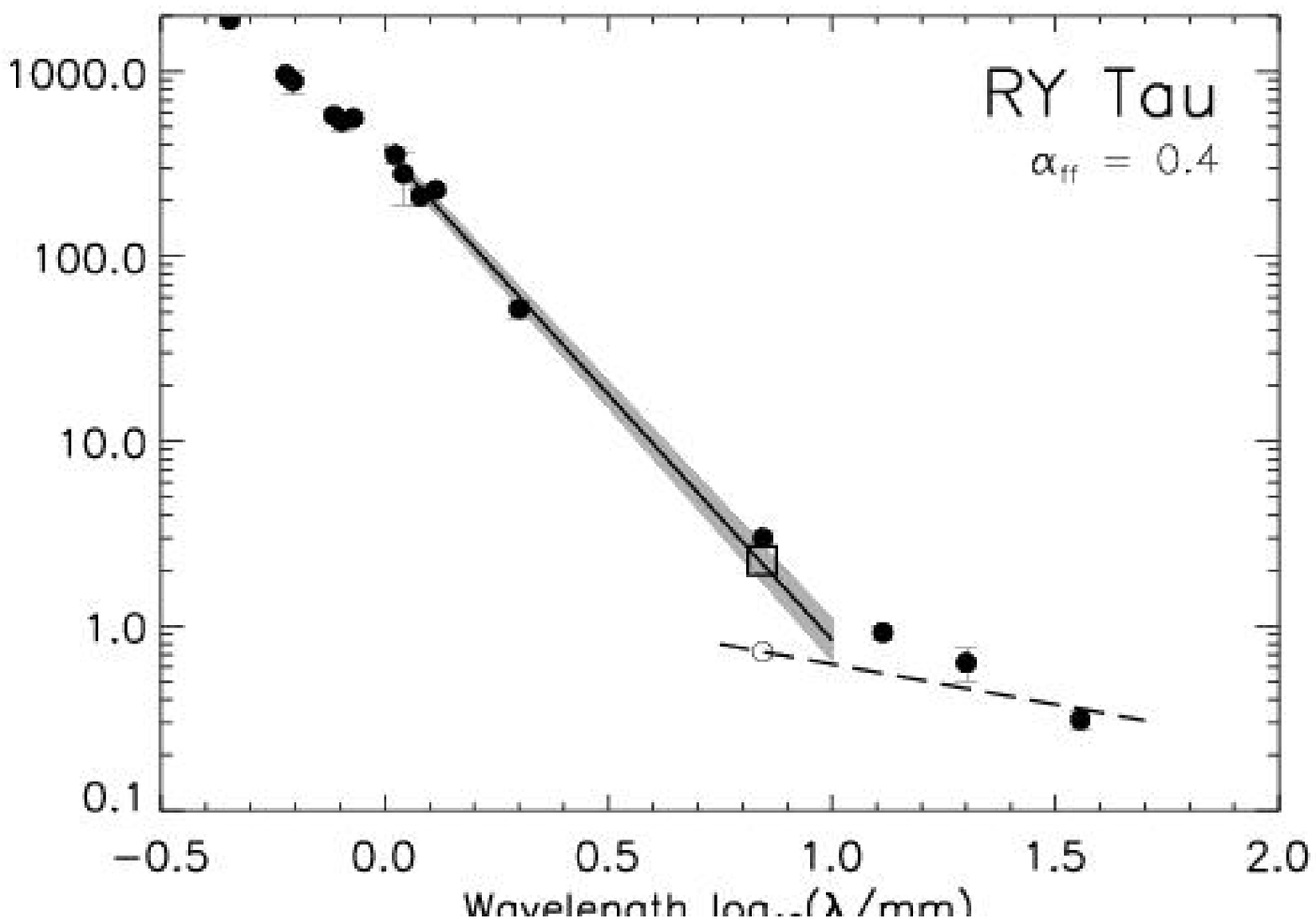}\figsp
  \includegraphics[width=0.45\textwidth]{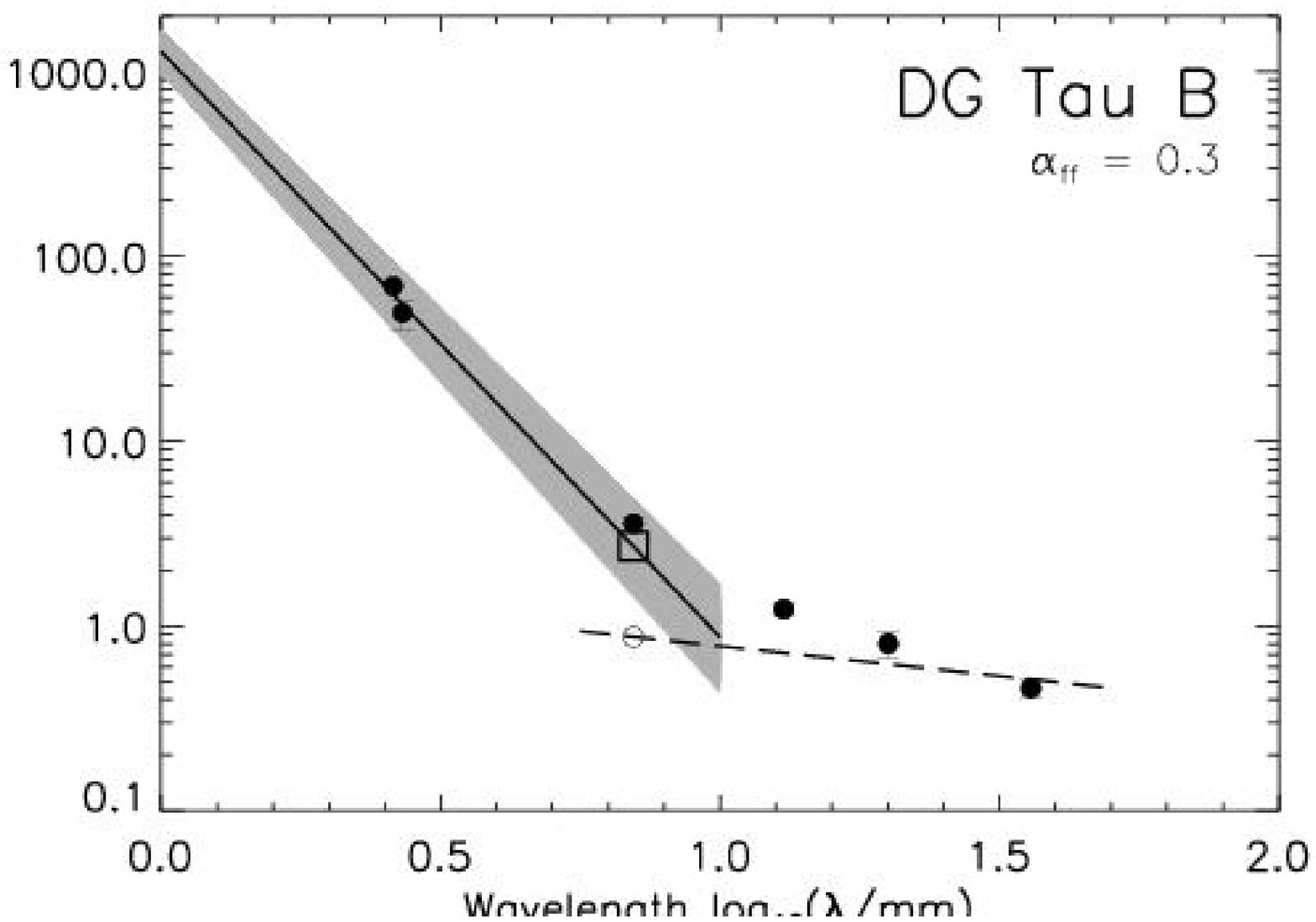}\figsp
  \includegraphics[width=0.45\textwidth]{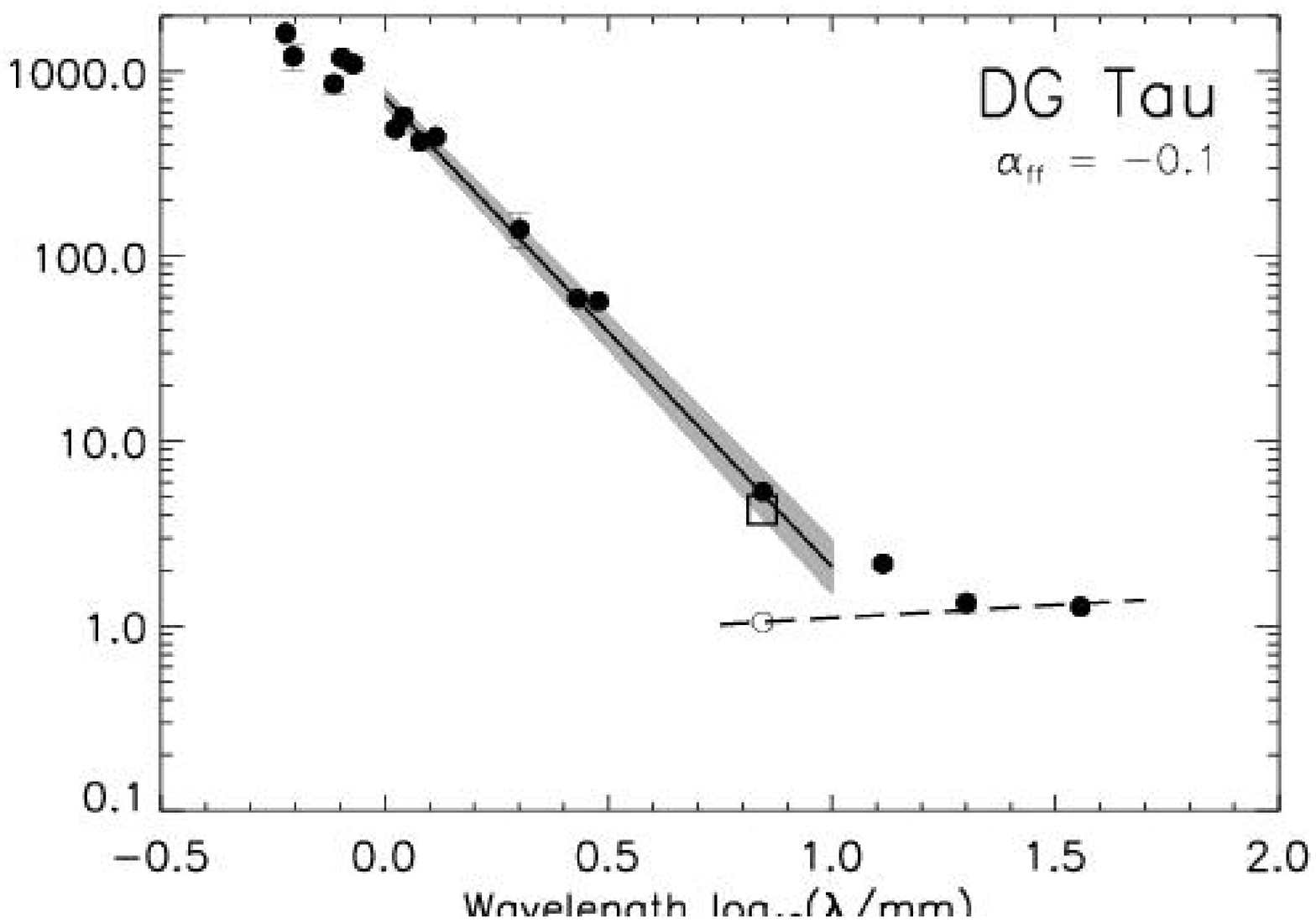}\figsp
  \includegraphics[width=0.45\textwidth]{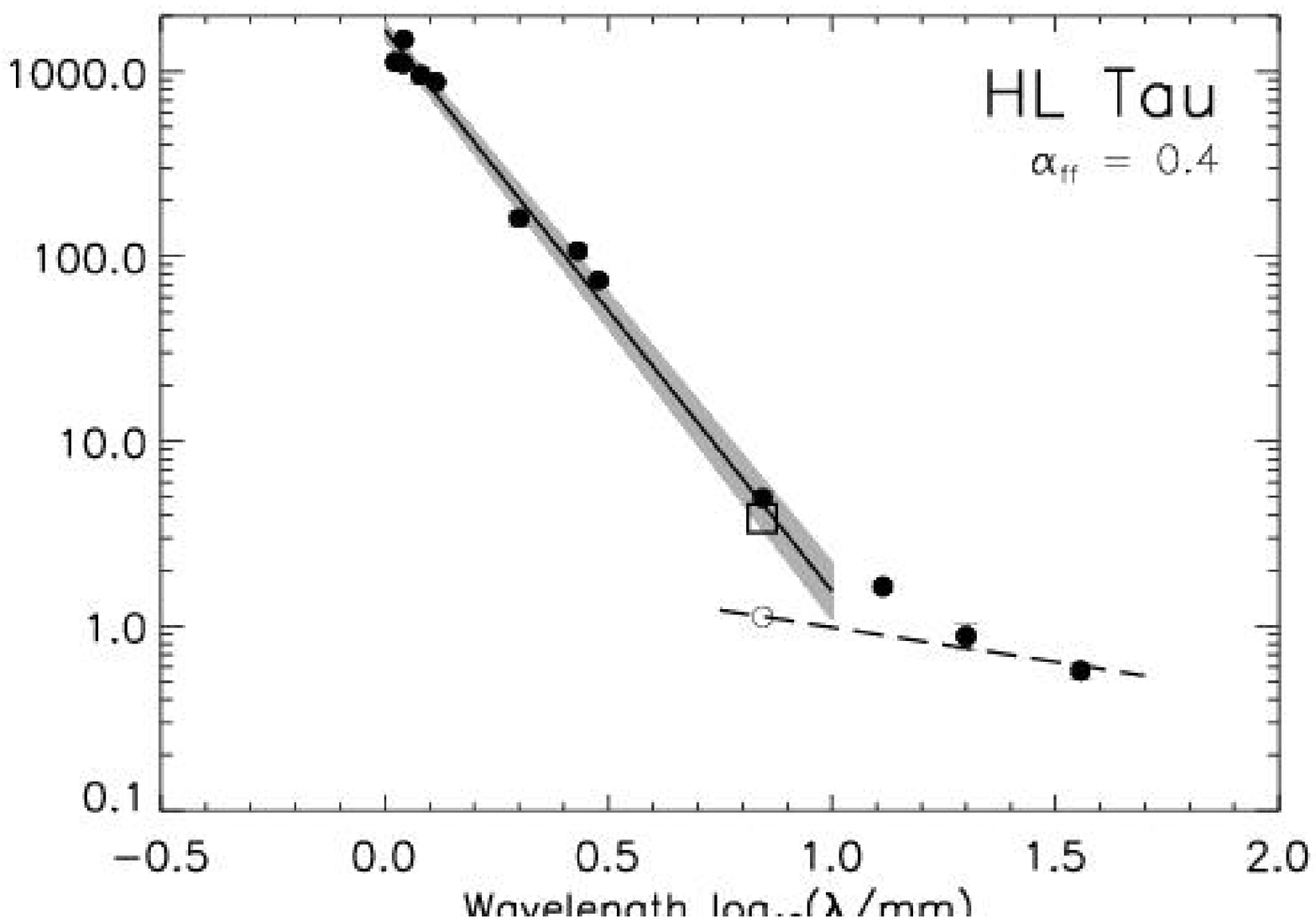}
  \caption{Free-free emission contribution to 7-mm flux densities. The dashed
    line represents the fit to the centimetre fluxes, assuming that 50\% of
    the 1.3-cm emission arises from free-free radiation. The square depicts 
    the corrected 7-mm flux after subtraction of the estimated free-free 
    contribution (shown as open circle). The thick line shows the power-law
    fit to the millimetre data; the shaded region indicates the uncertainty 
    range for the 1--7~mm slope. Data points shortwards of $\lambda$=7~mm were 
    compiled from the literature (Table \ref{fluxliterature}).}
  \label{ffcorr} 
\end{figure*}
%
%
%
%
At centimetre wavelengths the radio continuum of young stars is expected to be
dominated by free-free radiation (bremsstrahlung) from ionized winds or outflows.
There are, however, some observational indications that dust continuum emission 
may be a substantial source contributing to the centimetre fluxes of 
T~Tauri stars \citep{rodriguez94, wilner96, wilner05}. 

With the exception of \object{RY~Tau}, all other sources observed at
1.3~cm appear to be resolved, at least in one spatial direction. The 
position angles roughly match (within 20\deg) those derived from the 7-mm 
images. This finding supports the hypothesis that dust emission rather than an 
ionised wind is the main emission mechanism. For \object{HL~Tau} the position 
angles differ by $\sim$40\deg, probably as a result of the bright counter-jet. 

We measured the free-free spectrum for four sources by fitting a power law to 
the centimetre flux values (Figure~\ref{ffcorr}). The measured 2.0 and 3.6-cm 
fluxes are assumed to
originate from free-free radiation only. At higher frequencies, a reliable 
quantification of the corresponding contribution of thermal dust emission and 
free-free radiation is difficult; we assumed an equal mixture of dust and wind 
emission at 1.3~cm. The spectral index 
$\alpha_{\rm ff}$ ($S_\nu \propto \nu^{\alpha_{\rm ff}}$) was found to be in 
the range $-0.1$ to $+0.4$. 
A symmetric, ionized, and opaque wind with constant velocity has a spectral 
power index of 0.6 \citep{panagia75,wright75,olnon75}. $\alpha_{\rm ff}=-0.1$ 
is expected for totally transparent (optically thin) free-free emission 
\citep{mezger1967}.

For all four objects we 
extrapolated the free-free radio spectrum to $\lambda$=7~mm to 
estimate the contribution of wind emission to the 7-mm fluxes.
Subtracting the estimated free-free emission from the measured 7-mm 
value one obtains the thermal dust emission. We found that about 20\%
of the 7-mm emission originates from free-free radiation; 80\% are 
due to dust continuum emission. For sources where no centimetre data were 
available, we used these numbers to correct the 7-mm fluxes.

%
%
%
\subsection{Spectral indices and dust opacity indices}

The outer disk radii derived suggest that the emission is optically thin 
at 7~mm. One can therefore directly use the slope of the millimetre 
spectral energy distribution \mbox{($F_\nu \propto \nu^{\alpha}$)} 
to derive a dust opacity law \mbox{($\kappa_\nu \propto \nu^{\beta\,}$)}
\citep{beckwith91,testi03,natta04}.
Were the millimetre emission completely optically thin, i.e. without any 
contribution from the optically thick inner disk, one could simply read off 
the opacity index $\beta$ from the spectral index $\alpha$ via the relation 
\mbox{$\beta=\alpha-2$}. There is, however, a non-negligible contribution 
from the optically thick part of the disk to the measured fluxes. Knowing this 
contribution will enable an improved estimate for the emissivity index 
$\beta$.

The revised relation between the opacity (emissivity) index $\beta$ 
and the observed spectral index $\alpha$ is given by \mbox{$\beta \approx
  (\alpha-2)\times(1+\Delta)$}, where $\Delta$ is the ratio of optically thick
to optically thin disk emission given by
\mbox{$\Delta=-p \times \big\{ (2-q)\ln[(1-p/2)\,\bar{\tau}] \big\}^{-1} $}
\citep[Eq. 20 in][]{beckwith90}. Here $p$, $q$, and $\bar{\tau}$ are the 
power-law exponent of radial disk temperature, the power-law exponent of 
the surface density, and the average optical depth of the disk, respectively.

\begin{table*}
\caption{Spectral slopes derived from power-law fitting in the range 1--7~mm 
using literature values (Table~\ref{fluxliterature}) and our \mbox{7-mm} data 
(Col.~2). For objects without centimetre measurements, a 20\% contribution of 
free-free emission was assumed (denoted by brackets in Col.~3). 
Spectral slopes corrected for free-free radiation are listed in 
Col.~4. The estimated dust opacity indices $\beta^{\sim}$ (Col.~5) were corrected for 
optically depth (using $\Delta=0.2$), yielding the final $\beta$ values (Col.~6).}
\begin{center}
\begin{tabular}{l r@{\,\,$\pm$\,\,}l c r@{\,\,$\pm$\,\,}l  r@{\,\,$\pm$\,\,}l
  | r@{\,\,$\pm$\,\,}l |} 
\hline\\[-3mm]
Source     & \multicolumn{2}{c}{mm slope} & 
             \multicolumn{1}{c}{f--f contribution} &
             \multicolumn{2}{c}{f--f corrected} & 
	     \multicolumn{2}{c}{$\beta$ estimate} &
	     \multicolumn{2}{c}{Opacity index}\\ 
	   & \multicolumn{2}{c}{$\alpha_{\rm mm}$}      &
             \multicolumn{1}{c}{at 7 mm (\%)} &
             \multicolumn{2}{c}{$\alpha_{\rm corr}$} & 
	     \multicolumn{2}{c}{$\beta^{\sim}\!=\!\alpha_{\rm corr}\!-\!2$} &
	     \multicolumn{2}{c}{$\beta=(1+\Delta)\beta^{\sim}$}
\smallskip \\ \hline\hline\\[-2.5mm] 
\object{RY Tau}   & 2.51 & 0.10	 &  24  & 2.66 & 0.09  & 0.66 & 0.09  & \,\,\,\,\,\,\,0.8 & 0.1\\
\object{FT Tau}   & 2.62 & 0.17  & (20) & 2.76 & 0.24  & 0.76 & 0.24  & \,\,\,\,\,\,\,0.9 & 0.3\\
\object{DG Tau B} & 2.88 & 0.20  &  24  & 3.17 & 0.19  & 1.17 & 0.19  & \,\,\,\,\,\,\,1.4 & 0.2\\
\object{DG Tau}   & 2.45 & 0.09  &  20  & 2.54 & 0.11  & 0.54 & 0.11  & \,\,\,\,\,\,\,0.7 & 0.1\\
\object{HL Tau}   & 2.93 & 0.10  &  23  & 3.04 & 0.12  & 1.04 & 0.12  & \,\,\,\,\,\,\,1.3 & 0.1\\
\object{UZ Tau E} & 2.56 & 0.08  & (20) & 2.66 & 0.09  & 0.66 & 0.09  & \,\,\,\,\,\,\,0.8 & 0.1\\
\object{DL Tau}   & 2.73 & 0.11  & (20) & 2.82 & 0.14  & 0.82 & 0.14  & \,\,\,\,\,\,\,1.0 & 0.2\\
\object{CI Tau}   & 3.00 & 0.27  & (20) & 3.12 & 0.30  & 1.12 & 0.30  & \,\,\,\,\,\,\,1.3 & 0.4\\
\object{DO Tau}   & 2.29 & 0.06  & (20) & 2.38 & 0.07  & 0.38 & 0.07  & \,\,\,\,\,\,\,0.5 & 0.1\\
\object{GM Aur}   & 3.17 & 0.20  & (20) & 3.29 & 0.20  & 1.29 & 0.20  & \,\,\,\,\,\,\,1.6 & 0.2
\smallskip \\ 
\hline 
\end{tabular}
\label{spectralindex}
\end{center}
\end{table*}

For the surface-density profile we adopted a value of $p=1.5$ throughout. 
The temperature profile index $q$ is uniquely determined by the slope of the 
spectral energy distribution in the optically thick regime, given by the 
relation \mbox{$q = 2/(3-\alpha_{\rm IR})$}. We measured the spectral
slope at far-infrared wavelengths (see Table~\ref{fluxliterature}), 
and derived temperature indices in the range of $q \approx 0.5-0.7$. 
 Since $\Delta$ depends only weakly (logarithmically) on the average optical 
depth $\bar{\tau}$, an order-of-magnitude estimate for this quantity will 
suffice for our purposes. $\bar{\tau}$ is a function of the opacity, mass,
outer radius, and orientation of the disk \citep[Eq.~16 in][]{beckwith90}.
We estimate its approximate value by taking characteristic numbers for these 
parameters, i.\,e. disk masses from \cite{beckwith90} and disk radii from
Table~2, for a circumstellar disk seen at random orientation. We found that
$\bar{\tau}$ is of the order of $\sim$10$^{-2}$.  
After inserting all parameters in the correction formula we derive $\Delta 
\approx 0.2$. The optical-depth correction slightly increases the opacity 
indices. The final $\beta$ values are given in Table~\ref{spectralindex}.
%





\section{Discussion}
%
One can use the millimetre spectral slope of circumstellar disks around 
pre-main-sequence stars to gain information on the characteristic dust
grain size, provided the observed emission is largely optically thin.
The magnitude of the dust opacity index indicates whether dust aggregates 
are small or large compared to the observing wavelength. The $\beta$ values
are a robust measure of the characteristic grain size in the emitting
region. At 7~mm, we largely observe the thermal emission of cold dust in
the outer parts of the disk midplane.  
 
For dust grains with sizes of the order of a few tenths of a micron
\mbox{($\ll\lambda_{\rm obs} / 2\pi$)},
as present in the interstellar medium and in protostellar cores, 
the opacity index has been found to be \mbox{$\beta \approx 2$} 
\citep{hildebrand83,draine84,ossenkopf94}. For very large bodies 
\mbox{($\gg\lambda_{\rm obs} / 2\pi$)} that block radiation by virtue of 
their geometrical cross-section, the opacity is  
frequency-independent (grey opacity), \mbox{$\beta = 0$}. 

Particles of about the same size as the observing wavelength 
of $\lambda=7$~mm, i.\,e. pebble-sized particles, are expected to be in an 
intermediate regime \citep{beckwith00}. Strictly speaking, one has to include the 
refractive index \mbox{$m=n+{\mathrm i}k$} of the absorbing material. 
At millimetre wavelengths, the optical constants for silicate grains are 
$n \approx 3$ and $k\simeq 0$, thus $|m| \approx 3$ \citep{laor93, 
henning97,mutschke98}. Thus opacity indices $\beta\la 1$ suggest the
presence of dust particles with sizes \mbox{$\ga \lambda/ 2 \pi |m|$}, 
i.\,e. with millimetre dimensions (and possibly even larger).

\begin{figure*}
  \center\includegraphics[width=0.73\textwidth]{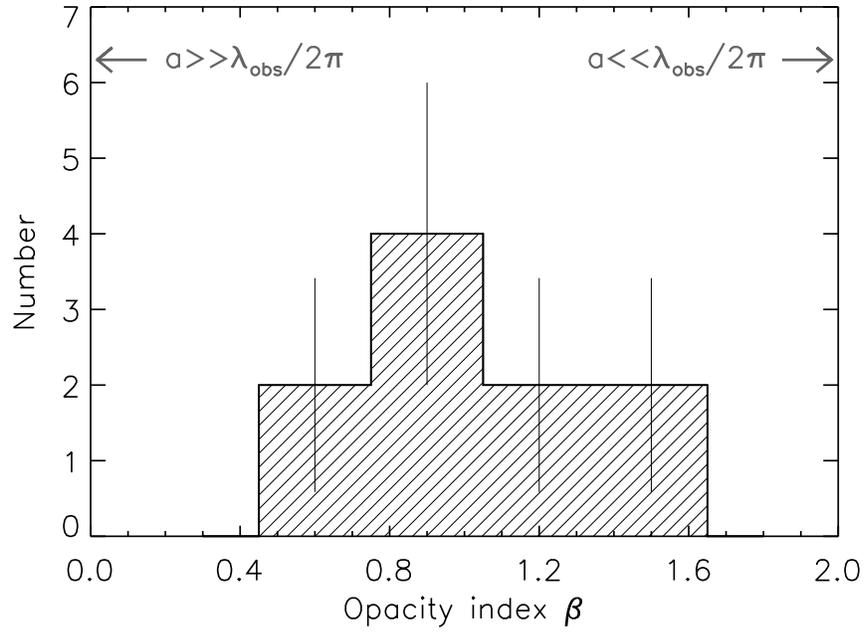}
  \caption{Distribution of dust opacity indices. ISM-sized dust
    grains have $\beta\simeq2$, while particles much larger than the
    observing wavelength $\lambda$=7~mm have a frequency-independent  
    opacity law ($\beta=0$). The populated intermediate regime is
    indicative for the presence of millimetre-sized dust aggregates. 
  }
  \label{betahist} 
\end{figure*}

After correcting for free-free radiation and optically thick emission, we 
determined dust opacity indices $\beta$ in the range \mbox{0.5--1.6} for 
10 sources where the circumstellar disk could be spatially resolved 
(Figure~\ref{betahist}). Six objects have $\beta$ value $\la 1$, a robust 
indication of agglomerated dust particles in the millimetre size regime. 
Four other sources have $\beta$ values between 1.3 and 1.6, lower than 
the dust opacity index $\beta\simeq 2$ of submicron-sized particles as 
found in the interstellar medium. $\beta$ values smaller than 2 may also 
be explained by grain properties 
like shape, composition, conductivity, porosity, and crystallinity of the
dust particles \citep{henning95}. Size, however, is the most important
parameter influencing dust opacity indices.
We deem unlikely the possibility that extreme dust compositions or particle
structures cause the observed low $\beta$ values \citep{koerner95,draine05}.

We underline the importance of proper free-free correction of the
millimetre slopes, without which the opacity indices would be systematically
smaller. The detection of emission at centrimetric wavelengths remains the
only safe method to estimate the contamination of free-free radiation to
the millimetre emission observed \citep{testi01,natta04}.

\section{Conclusion}
%

We presented continuum observations at 7~mm of a large sample of
low-mass pre-main-sequence stars located in the Taurus-Auriga
star-forming region with ages between approximately 100~kyr and 3~Myr.
Circumstellar material could be spatially resolved for 10 sources, 
allowing us to derive the size of their circumstellar disk.

The deconvolved disk sizes (100--200~AU) effectively rule out the 
possibility that the 7~mm emission originates from small, optically 
thick disks containing submicron-sized dust particles. Knowing that the
millimetre emission is mostly optically thin, we could directly
translate the spectral indices $\alpha_{\rm mm}$ measured in the
millimetre part of the spectral energy distribution into the
frequency dependence of the dust opacity.

The power-law index $\beta$ is indicative for the characteristic 
size of the radiating dust particles. From our 7~mm observation and other 
millimetre fluxes compiled from the literature, we found $\beta$ values 
covering the range \mbox{0.5--1.6}. More than half of the objects where 
the disk surrounding the star could be spatially resolved show clear 
evidence for the presence of millimetre-sized dust particles, demonstrating 
the operation of dust grain growth processes in disks around T~Tauri stars.
The remaining sources may not require large dust aggregates and are 
consistent with submicron-sized dust as found in the interstellar medium. 

Together with similar results for \object{TW~Hya}, \object{CQ~Tau},
and two Herbig Ae stars (\object{HD~34282}, \object{HD~163296})
\citep{calvet02,testi03,natta04}, a picture of grain growth to
pebble-sized particles in disks around low- to intermediate-mass 
pre-main-sequence stars is emerging. Further investigations of
grain growth with existing and upcoming millimetre interferometers are 
needed to study where and how fast the building blocks of planetesimals
are formed.

Millimetre observations trace the cold outer disk midplane where most of 
the disk mass is locked up. The grain-size indicator $\beta$ therefore
only conveys information on that disk region. The surface layer of the 
inner disk is accessible through mid-infrared spectroscopy.
The comparison of dust-size indicators from millimetre interferometry 
and mid-infrared spectroscopy, sampling different regions of circumstellar 
disks around T~Tauri stars, may yield valuable insights into the physical 
processes that are thought to control the formation of planetesimals and the 
evolution of the circumstellar disk. Establishing the characteristic dust
grain sizes at different locations in circumstellar disks holds the key to a 
better understanding of planet formation and disk evolution.

\begin{acknowledgements}
\mbox{J.\,R.} would like to thank Andreas Brunthaler (Joint Institute for VLBI
in Europe, JIVE) and Hendrik Linz (MPIA, Heidelberg) for their helpful 
introduction into the world of $\mathcal{AIPS}$. We also thank Ralf Launhardt 
for a critical reading of the manuscript. Partial support for
\mbox{D.\,J.\,W.} for this work was provided by NASA Origins of Solar System
Program Grant NAG5-11777. 
  
 This research has made extensive use of the SIMBAD database and VizieR 
catalogue service, operated at CDS, Strasbourg, France.
\end{acknowledgements}

\begin{landscape}
\vspace*{2mm}
\begin{center}
%
%
\begin{table}[h]
\caption{Millimetre measurements and IRAS photometry compiled from the literature. The columns~\beam\ list 
the HPBW; note that the IRAS PSF sizes are given in arcmin.} 
\begin{tabular}{ l || llcc | llcc | llcc | llcc | llcc }
\hline\\[-3mm]
$\;\;\;\,\lambda$ & 
\multicolumn{1}{c}{Flux}  & 
\multicolumn{1}{c}{Error} & 
\multicolumn{1}{c}{\beam} & 
Ref. &
\multicolumn{1}{c}{Flux}  & 
\multicolumn{1}{c}{Error} & 
\multicolumn{1}{c}{\beam} & 
Ref. & 
\multicolumn{1}{c}{Flux}  & 
\multicolumn{1}{c}{Error} & 
\multicolumn{1}{c}{\beam} & 
Ref. & 
\multicolumn{1}{c}{Flux}  & 
\multicolumn{1}{c}{Error} & 
\multicolumn{1}{c}{\beam} & 
Ref. & 
\multicolumn{1}{c}{Flux}  & 
\multicolumn{1}{c}{Error} & 
\multicolumn{1}{c}{\beam} & 
Ref. \\
(mm)  & 
\multicolumn{1}{c}{(Jy)}     & 
\multicolumn{1}{c}{(Jy)}     & 
\multicolumn{1}{c}{(\arcsec)}&   
&
\multicolumn{1}{c}{(Jy)}     &
\multicolumn{1}{c}{(Jy)}     & 
\multicolumn{1}{c}{(\arcsec)}&       
& 
\multicolumn{1}{c}{(Jy)}     &
\multicolumn{1}{c}{(Jy)}     & 
\multicolumn{1}{c}{(\arcsec)}&      
&
\multicolumn{1}{c}{(Jy)}     &
\multicolumn{1}{c}{(Jy)}     & 
\multicolumn{1}{c}{(\arcsec)}&      
&
\multicolumn{1}{c}{(Jy)}     & 
\multicolumn{1}{c}{(Jy)}     &
\multicolumn{1}{c}{(\arcsec)}&
\smallskip \\ \hline\hline\\[-2.5mm]  
%
&\multicolumn{4}{c}{\bf{RY Tau}}  & 
 \multicolumn{4}{c}{\bf{FT Tau}}  & 
 \multicolumn{4}{c}{\bf{DG Tau B}}& 
 \multicolumn{4}{c}{\bf{DG Tau}}  & 
 \multicolumn{4}{c}{\bf{HL Tau}}   \\

0.012 & 17.74& 0.027& 1\am& (a)  & 0.36 & 0.045& 1\am& (a)  &      &      &     &      &      &      &     &      &  9.74& 0.031& 1\am& (a) \\
0.025 & 26.48& 0.054& 1\am& (a)  & 0.57 & 0.037& 1\am& (a)  &      &      &     &      & 30.39& 0.065& 1\am& (a)  & 31.18& 0.073& 1\am& (a) \\
0.060 & 18.91& 0.066& 2\am& (a)  & 0.82 & 0.041& 2\am& (a)  &      &      &     &      & 31.28& 0.039& 2\am& (a)  & 76.26& 0.810& 2\am& (a) \\
0.100 & 13.50& 2.501& 4\am& (a)  &      &      &     &      &      &      &     &      & 39.86& 0.840& 4\am& (a)  & 77.95& 1.394& 4\am& (a) \\
1.056 & 0.354& 0.036& 22  & (b)  &      &      &     &      &      &      &     &      & 0.489& 0.033& 22  & (b)  &  1.13& 0.02 & 22  & (b) \\
1.1   & 0.28 & 0.09 & 18.5& (f)  &      &      &     &      &      &      &     &      & 0.57 & 0.07 & 18.5& (f)  &  1.11& 0.02 & 18  & (k) \\
1.2   & 0.212& 0.021& 12  & (g)  &      &      &     &      &      &      &     &      & 0.420& 0.042& 12  & (g)  & 0.961& 0.096& 12  & (g) \\
1.3   & 0.229& 0.017& 11  & (c)  & 0.130& 0.014& 11  & (c)  &      &      &     &      & 0.443& 0.020& 11  & (c)  & 0.879& 0.019& 11  & (c) \\
2.0   & 0.052& 0.006&  5  & (d)  &      &      &     &      &      &      &     &      & 0.14 & 0.03 & 27.5& (f)  & 0.161& 0.017& 5   & (d) \\
2.6   &      &      &     &      &      &      &     &      &0.0691&0.0017&  4  & (h)  &      &      &     &      &      &      &     &     \\
2.7   &      &      &     &      & 0.025&0.0022&  3  & (e)  &0.0494&0.0088& 0.7 & (i)  & 0.059&0.0020& 3   & (e)  &0.1069&0.0078& 0.6 & (i) \\
3.0   &      &      &     &      &      &      &     &      &      &      &     &      & 0.057&0.0040& 6   & (j)  &0.074 &0.0040& 3   & (j) \\

\hline\\[-2.5mm] 

&\multicolumn{4}{c}{\bf{UZ Tau E}} &
 \multicolumn{4}{c}{\bf{DL Tau}}   &
 \multicolumn{4}{c}{\bf{CI Tau}}   &
 \multicolumn{4}{c}{\bf{DO Tau}}   &
 \multicolumn{4}{c}{\bf{GM Aur}}   \\

0.012 &  1.38& 0.031& 1\am& (a)  & 0.97 & 0.034& 1\am& (a)  & 0.78 & 0.025& 1\am& (a)  &      &      &     &      & 0.25 & 0.031& 1\am& (a) \\
0.025 &  1.76& 0.043& 1\am& (a)  & 1.32 & 0.045& 1\am& (a)  & 1.30 & 0.046& 1\am& (a)  & 4.07 & 0.041& 1\am& (a)  & 1.07 & 0.042& 1\am& (a) \\
0.060 &  2.37& 0.069& 2\am& (a)  & 1.39 & 0.084& 2\am& (a)  & 2.15 & 0.071& 2\am& (a)  & 6.33 & 0.106& 2\am& (a)  & 3.08 & 0.112& 2\am& (a) \\
0.100 &      &      &     &      &      &      &     &      &      &      &     &      & 8.55 & 1.483& 4\am& (a)  &      &      &     &     \\  
1.056 & 0.213& 0.039& 22  & (b)  & 0.273& 0.037& 22  & (b)  & 0.217& 0.050& 22  & (b)  & 0.194& 0.032& 22  & (b)  &      &      &     &     \\
1.1   &      &      &     &      & 0.26 & 0.03 & 18.5& (f)  &      &      &     &      & 0.18 & 0.02 & 18.5& (f)  & 0.38 & 0.03 & 18  & (k) \\
1.2   & 0.170& 0.017& 12  & (g)  & 0.198& 0.020& 12  & (g)  & 0.177& 0.018& 12  & (g)  & 0.145& 0.015& 12  & (g)  &      &      &     &     \\
1.3   & 0.137& 0.028& 1.1 & (l)  & 0.230& 0.014& 11  & (c)  & 0.190& 0.017& 11  & (c)  &0.1375& 0.032&  3  & (m)  & 0.253& 0.012& 11  & (c) \\
1.35  &      &      &     &      &      &      &     &      &      &      &     &      &0.0986& 0.024&  4  & (m)  &      &      &     &     \\
2.0   &      &      &     &      & 0.12 & 0.03 & 27.5& (f)  &      &      &     &      & 0.037& 0.005&  5  & (d)  & 0.037& 0.004& 5   & (d) \\
2.7   &0.0255&0.0016& 3   & (e)  &0.0296&0.0020&  3  & (e)  &0.0310&0.0033& 3   & (e)  &0.0255&0.0027&  3  & (e)  &0.0277&0.0022& 3   & (e) \\
3.0   &      &      &     &      & 0.023&0.0040&  8  & (j)  &      &      &     &      &      &      &     &      &      &      &     &     \\
3.1   & 0.014&0.0031& 2.1 & (l)  &      &      &     &      &      &      &     &      &      &      &     &      &      &      &     &     \\
3.4   &      &      &     &      &      &      &     &      &      &      &     &      &0.0142&0.0007&  1  & (m)  &      &      &     &     \\

\hline
\end{tabular}
\vspace*{2mm}
\flushleft References --- 
        (a)~\citet{weaver92};
        (b)~\citet{beckwith91};
	(c)~\citet{beckwith90};
	(d)~\citet{kitamura02};
	(e)~\citet{dutrey96};
	(f)~\citet{mannings94};
	(g)~\citet{altenhoff94};
	(h)~\citet{mitchell97};
	(i)~\citet{looney00};
	(j)~\citet{ohashi96};
	(k)~\citet{adams90};
	(l)~\citet{jensen96};
	(m)~\citet{koerner95}.
\label{fluxliterature}
\end{table}
\end{center}
\end{landscape}

\bibliographystyle{aa}       
\bibliography{refs}  

\if@referee
\else
\fi

\end{document}